\documentclass[twocolumn,english,aps,prb,showpacs]{revtex4}
\usepackage[T1]{fontenc}
\usepackage[latin1]{inputenc}
\usepackage{graphicx}
\usepackage{amssymb}

\makeatletter

\providecommand{\LyX}{L\kern-.1667em\lower.25em\hbox{Y}\kern-.125emX\@}

\usepackage{babel}
\makeatother
\begin{document}

\newcommand{\kt}[1]{\left|#1\right\rangle }

\newcommand{\br}[1]{\left\langle #1\right|}

\newcommand{\ph}{\hat{\phi }}

\newcommand{\pd}{\hat{\psi }^{\dagger }}

\newcommand{\ps}{\hat{\psi }}

\title{Dephasing in sequential tunneling through a double-dot interferometer
{\small }\\
}

\author{Florian Marquardt and C. Bruder}

\affiliation{Departement Physik und Astronomie, Universität Basel, Klingelbergstr.
82, CH-4056 Basel}

\date{March 19th, 2003}

\email{Florian.Marquardt@unibas.ch}

\begin{abstract}
We analyze dephasing in a model system where electrons tunnel sequentially
through a symmetric interference setup consisting of two single-level
quantum dots. Depending on the phase difference between the two tunneling
paths, this may result in perfect destructive interference. However,
if the dots are coupled to a bath, it may act as a which-way detector,
leading to partial suppression of the phase-coherence and the reappearance
of a finite tunneling current. In our approach, the tunneling is treated
in leading order whereas coupling to the bath is kept to all orders
(using $P(E)$ theory). We discuss the influence of different bath
spectra on the visibility of the interference pattern, including the
distinction between {}``mere renormalization effects'' and {}``true
dephasing''.
\end{abstract}


\maketitle

\section{Introduction}

The destruction of quantum-mechanical phase coherence due to coupling
of a system to an irreversible bath is a subject important not only
because of its connection to fundamental issues (the quantum measurement
process and the quantum-classical transition) but also because of
its role in the suppression of phenomena resulting from quantum interference
effects, such as those studied in mesoscopic physics (including Aharonov-Bohm
interference, weak localization and universal conductance fluctuations).
Recently, the field of mesoscopic physics in particular has seen a
revival of interest in these questions, due to surprising experimental
findings\cite{mohantywebb} concerning a possible saturation of the
weak-localization dephasing rate at low temperatures, that have not
yet been explained convincingly. Apart from investigations dealing
directly with the problem of weak localization in a disordered system
of interacting electrons, several toy models have been analyzed\cite{cedraschi,buettrevb,GZ_PB,OurABring,NagaevBuettikerHO,Guinea,GSZ_modelsLowTDeph,ImryInexistenceZPDP}
to answer the question whether decoherence at zero temperature is
possible at all, contrary to the expectations based on perturbation
theory. One of the difficulties faced by models involving discrete
levels consists in the fact that destruction of phase coherence for
a superposition of excited states of finite excitation energy is perfectly
possible even at zero temperature, due to spontaneous emission of
energy into the bath. It is only in the zero-frequency limit of the
linear response in a system with a continuous spectrum (relevant for
weak-localization and other equilibrium transport experiments) that
perturbation theory suggests in general a vanishing dephasing rate,
because then the perturbation does not supply energy to the system,
such that at $T=0$ the system is not able to leave a trace in the
bath, which is considered to be the prerequisite for decoherence.

Some questions of interest concerning dephasing, especially in connection
with mesoscopic systems and low temperatures, are the following ones:
How reliable is the simple classical picture of a phase being randomized
by fluctuating external noise\cite{sai}? In particular, what is the
meaning of the zero-point fluctuations of the bath in this picture,
as opposed to the thermal fluctuations dominating at frequencies lower
than the temperature? When do the former lead to {}``mere renormalization
effects'' and how is it possible to distinguish these from {}``true''
dephasing? Under which circumstances is the suppression of off-diagonal
terms in the reduced system density matrix itself already a good indicator
of dephasing? How reliable are simple arguments based on Golden Rule
and energy conservation, related to the connection between dephasing
and the trace left in the bath by the particle ({}``which-way''
detection)? When does perturbation theory fail qualitatively, what
is the influence of non-Markoffian behaviour? How does the dephasing
rate depend on the energy supplied by an external perturbation (frequencies
excited in linear response, bias voltage applied in a transport measurement)?
What is the influence of the Pauli principle in a system of degenerate
fermions? How strong are the qualitative differences in behaviour
resulting from different bath spectra?

In this work, we will present a model that is able to give insights
into most of these questions. 

\begin{figure}
\begin{center}\includegraphics[  height=5cm]{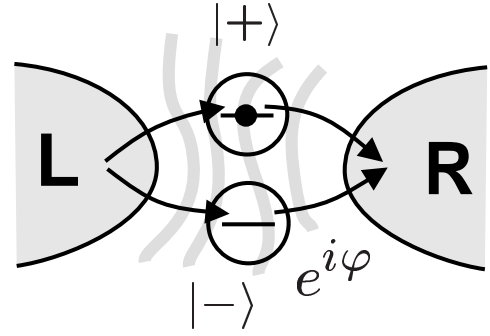}\end{center}

\caption{\label{DDdoubleslitsetup}The double-dot {}``double-slit'' setup,
with a fixed phase difference $\varphi $ between the two paths and
under the influence of a fluctuating environment.}
\end{figure}

Our model represents a kind of mesoscopic double-slit setup. It consists
of two single-level quantum dots which are tunnel-coupled to two leads,
with a possible phase difference between the two interfering paths
(see Fig. \ref{DDdoubleslitsetup}). Due to destructive interference
(at $\varphi =\pi $), the tunneling current may be suppressed completely,
provided the two dot-levels are degenerate and the setup is symmetric
in the two interfering paths. Coupling the dots to a bath may partly
destroy the phase coherence and re-enable the electrons to go through
the setup. For a symmetric setup, with equal coupling strength between
the bath and each of the two dots, mere renormalization effects will
not be able to lift the destructive interference in this way. Thus,
a finite tunneling current may be taken as a genuine sign of dephasing.
This criterion for dephasing has been employed before in a model of
dephasing due to spin-flip transitions in first-order tunneling transport
through one or two dots\cite{koeniggefen}, as well as for cotunneling
through an Aharonov-Bohm ring coupled to a fluctuating magnetic flux\cite{OurABring}. 

The influence of phonons on sequential tunneling through two quantum
dots \emph{in series} has been studied experimentally in Ref.~\onlinecite{KouwenhouwenDD}.
There, inelastic transitions induced by piezoelectric coupling to
acoustic phonons in GaAs have been essential for obtaining a finite
current through the two off-resonant dot levels. This kind of setup
has been analyzed theoretically in Refs.~\onlinecite{StoofNazarov,brune,ziegler,BrandesKramer,aguado,Debald,KeilSchoellerRTRG}.
On the other hand, we will be analyzing tunneling through two dots
placed \emph{in parallel}. Early theoretical investigations of this
problem (without a fluctuating environment) include Refs.~\onlinecite{akera,shahbazyan}.
Recently, a parallel-dot tunneling setup has been realized experimentally
in Ref.~\onlinecite{KotthausDD}, with an emphasis on spectroscopy of the
{}``molecular states'' of the double-dot system (with inter-dot
tunneling present). In our model of an interference setup, we choose
to describe a situation without tunneling between the dots (but with
Coulomb-repulsion). In addition, we want to concentrate on interference
effects in the orbital motion and therefore consider the case of spin-polarized
transport. This model - in the absence of a fluctuating environment
- has been investigated previously in Ref.~\onlinecite{BoeseDD}. Other
recent theoretical works concerning tunneling through dots in a parallel
geometry have mostly investigated spin and Kondo physics \cite{LossShenjaDD,mahnsoo,BoeseHofstetterSchoeller},
but also dephasing by spin-flip transitions\cite{koeniggefen}. Some
works have treated the influence of phonons in tunneling interference
structures\cite{HauleBonca,key-2}, but no systematic discussion of
dephasing and the visibility of the interference pattern has been
given. Some while ago, dephasing by \emph{nonequilibrium} current
noise has been investigated experimentally\cite{key-1} and theoretically\cite{aleinerwhichpath}
in a setup with a single quantum-dot placed into one arm of an Aharonov-Bohm
interferometer.

Our analysis of dephasing in sequential tunneling through a double-dot
will take into account the system-bath coupling exactly, while we
treat the tunnel-coupling only in leading order. The presence of the
Fermi sea in the leads introduces some aspects related to the Pauli
principle and to the behaviour of systems with a continuous spectrum
that cannot be analyzed in simpler models of dephasing in discrete
systems coupled to a bath.

The work is organized as follows: After setting up the model (Sec.
\ref{sec:The-model}), we will present a qualitative discussion of
its main features (\ref{sec:Qualitative-discussion}). In particular,
we will discuss the relation between entanglement, dephasing and renormalization
effects. Subsequently, we derive a general formula for the tunneling
decay rate of an electron that has been placed on the two dots in
a symmetric superposition of states (Sec. \ref{sec:Decay-rate-and}).
This is done by building on the concepts of the $P(E)$ theory of
tunneling in a dissipative environment \cite{ingoldhabil,SchoenPE}.
Following this, we will evaluate the dependence of the tunneling rate
on the bias voltage and the bath spectra (Sec. \ref{sec:Evaluation-for-different}).
We will interpret the results in terms of {}``renormalization effects''
and {}``true dephasing'' (Sec. \ref{sec:Discussion-of-the}). Building
on these sections, we will finally derive a master equation for the
case of weak tunnel coupling (Sec. \ref{DDseqTunnelingSection}),
which allows us to calculate the sequential tunneling current as a
function of bias voltage, temperature, and phase difference (Sec.
\ref{sec:Evaluation-of-the}).

The most important results derived in this work are the following:
Equation (\ref{DDGammabasic}) is the general expression for the phase-dependent
tunneling decay rate in presence of the fluctuating environment. It
forms the basic input for the master equation (Eqs. (\ref{DDrhoppeq})-(\ref{DDrhopmEq})),
that describes sequential tunneling through the double-dot, where
the resulting current can be obtained from Eq. (\ref{DDcurrentexpression}).
The visibility of the interference pattern, which is defined by the
phase-dependence of the current, is given in Eq. (\ref{DDvisibilityI}).
It is connected with the visibility obtained from the phase-dependence
of the tunneling rate itself (Eqs. (\ref{eq:visDef}), (\ref{eq:visGamm})).

\section{The model}

\label{sec:The-model}We consider a Hamiltonian describing two degenerate
single-level quantum dots, with respective single-particle states
$\left|+\right\rangle $ and $\left|-\right\rangle $ (spin is excluded
for simplicity, since we are interested in dephasing of the electronic
motion). Each of them is tunnel-coupled to two electrodes (with the
same strength for both dots), but involving a possible phase difference
between the tunnel amplitudes (see Fig. \ref{DDdoubleslitsetup}).
In addition, the potential difference between the two dots is given
by a fluctuating field $\hat{F}$, whose dynamics is derived from
a linear bath. It represents the fluctuations due to phonons or Nyquist
noise. The system-bath coupling strength is taken to be the same for
both dots, while the sign is opposite, such that the bath can distinguish
between an electron being on dot $\left|+\right\rangle $ or $\left|-\right\rangle $:

\begin{eqnarray}
\hat{H} & = & \epsilon (\hat{n}_{+}+\hat{n}_{-})+\hat{F}(\hat{n}_{+}-\hat{n}_{-})+U\hat{n}_{+}\hat{n}_{-}+\nonumber \\
 &  & \hat{H}_{L}+\hat{H}_{R}+\hat{H}_{B}+\hat{V}\label{DDhamiltonian}
\end{eqnarray}

Here $\hat{n}_{\pm }$ are the particle numbers on the two dots (equal
to $0$ or $1$). The bath Hamiltonian $\hat{H}_{B}$ describes a
set of uncoupled harmonic oscillators. It governs the dynamics of
the fluctuating potential $\hat{F}$, which is assumed to be linear
in the oscillator coordinates. The coupling between electron and bath
is of the form of the {}``independent boson model''\cite{mahan}.
For the case of exactly one electron on the double-dot, and in the
absence of tunneling, it corresponds to a spin-boson model with {}``diagonal
coupling''. In this model, no transition between different levels
is brought about by the bath, such that pure dephasing results. $U$
denotes the Coulomb repulsion energy, which we will take to be so
large that double-occupancy is forbidden. Note that the degeneracy
of the two dot-levels is important in the following: It is necessary
to ensure complete destructive interference at $\varphi =\pi $ (compare
also the discussion in Sec. \ref{DDseqTunnelingSection}). 

The terms $\hat{H}_{L}$ and $\hat{H}_{R}$ contain the energies of
the electrons in the left and right reservoirs: 

\begin{equation}
\hat{H}_{L(R)}=\sum _{k}\epsilon _{k}\hat{a}_{L(R)k}^{\dagger }\hat{a}_{L(R)k}\, .\end{equation}
The tunneling between the dots and the leads is described by $\hat{V}=\hat{V}_{L}+\hat{V}_{R}$,
with

\begin{equation}
\hat{V}_{R}=\sum _{k}t_{k}^{R}\hat{a}_{Rk}^{\dagger }(\hat{d}_{+}+e^{i\varphi }\hat{d}_{-})+h.c.\label{DDtunnelHam}\end{equation}

for the right junction, and 

\begin{equation}
\hat{V}_{L}=\sum _{k}t_{k}^{L}\hat{a}_{Lk}^{\dagger }(\hat{d}_{+}+\hat{d}_{-})+h.c.\label{eq:DDtunnelL}\end{equation}

for the left junction.

Here $\hat{d}_{\pm }$ are the annihilation operators for the two
dots ($\hat{n}_{\pm }=\hat{d}_{\pm }^{\dagger }\hat{d}_{\pm }$) and
the phase-factor of $e^{i\varphi }$ controls the interference between
tunneling events along either the upper or lower path. The tunneling
phase difference might be thought of as arising due to the Aharonov-Bohm
phase from a magnetic flux penetrating the region between the quantum
dots. 

Note that the tunneling matrix elements $t_{k}^{R(L)}$ are assumed
not to depend on the dot state $\kt{+}$ or $\kt{-}$ in our model.
This means that the dots are close enough such that they couple to
the same point on the lead electrodes, to within less than a Fermi
wavelength. Obviously there could be no appreciable interference effect
if the dots were separated by some larger distance (in which case
the $k$-dependence of matrix elements would be different for the
two states). The same idealized assumption underlies several similar
models (see, e.g., Refs.~\onlinecite{koeniggefen,BoeseDD,BoeseHofstetterSchoeller}).
The effect of an arbitrary dot separation has been discussed in some
detail in Ref.~\onlinecite{shahbazyan}. 

The present model, without the bath, has been analyzed previously
in Ref.~\onlinecite{BoeseDD} (see also Sec. IV.C of Ref.~\onlinecite{koeniggefen}).
There, an orbital type of Kondo effect was found in equilibrium, for
$\varphi =\pi $, when the level energy was below the chemical potential.
This arises because at $\varphi =\pi $ there are two states of the
double-dot that couple only to the left and the right lead, respectively
(denoted by $\kt{e}$ and $\kt{o}$ in the following). These degenerate
states form the pseudospin responsible for the Kondo effect. However,
that mechanism will be irrelevant for our analysis, as we consider
the transport situation where the (renormalized) level energy lies
between the chemical potentials of the left and the right lead. Therefore,
the degeneracy is effectively lifted by the bias voltage (which will
be assumed to be much larger than the tunneling rate), and only the
state coupling to the left lead would be occupied at $\varphi =\pi $.

\section{Qualitative discussion}

\label{sec:Qualitative-discussion}In this and the following three
sections, we first analyze the escape of a single electron into the
right lead, where the electron is assumed to start out in a symmetric
superposition of the two dot levels, which has been formed by an electron
tunneling onto the dots from the left lead. In the situation without
any bath, this is the state $\left|e\right\rangle \equiv (\left|+\right\rangle +\left|-\right\rangle )/\sqrt{2}$. 

Without dephasing, the tunneling decay out of state $\kt{e}$ is made
impossible in the case of perfect destructive interference at $\varphi =\pi $,
while maximal constructive interference is present for $\varphi =0$.
It should be noted that the attribution of the phase factor to one
of the tunnel couplings represents a certain choice of gauge, which
affects the wave functions in the following discussion but none of
the physically observable quantities that are derived as a result
of the master equation in Section \ref{DDseqTunnelingSection}.

\begin{figure}
\begin{center}\includegraphics[  height=5cm]{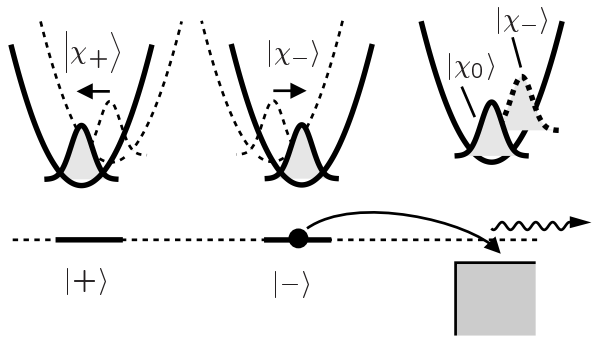}\end{center}

\caption{\label{DDevpic}The ground state $\kt{\chi _{+}}$ ($\kt{\chi _{-}}$)
which the bath assumes in the presence of an electron on dot $\kt{+}$
($\kt{-}$), shown schematically for a single bath oscillator (see
main text). After the electron has tunneled into the lead, $\kt{\chi _{-}}$
becomes a superposition of excited states (dashed), while the state
$\kt{\chi _{0}}$ represents the ground state of the bath in the new
potential.}
\end{figure}

For simplicity, we will assume a zero-temperature situation throughout
the following qualitative discussion, with a bias $eV>0$ applied
between the two dots and the lead in such a way that the electron
is allowed to tunnel into the lead (see Fig. \ref{DDevpic}). In addition,
since we will describe the tunneling decay to the right, we will only
consider the coupling $\hat{V}_{R}$ to the right lead in this section
and drop the index $R$ for now.

Without the bath and for perfect constructive interference ($\varphi =0$),
the tunneling decay rate $\Gamma $ will take on its maximum value
of $2\Gamma _{0}$, with

\begin{equation}
\Gamma _{0}\equiv 2\pi D\left\langle \left|t_{k}\right|^{2}\right\rangle \, ,\label{DDbaretunnelrate}\end{equation}

where $D$ is the lead density of states at the Fermi energy, $\left\langle \left|t_{k}\right|^{2}\right\rangle $
is the angular average of $\left|t_{k}\right|^{2}$ at this energy.
The bias voltage $V$ does not enter in this case, as long as it is
positive (permitting decay). For $\varphi =\pi $, $\Gamma $ vanishes
due to perfect destructive interference. In general, we have:

\begin{equation}
\Gamma =\Gamma _{0}(1+\cos \varphi )\, .\label{eq:nobath}\end{equation}

If the bath is included in the description, the following happens:

First of all, the energy of a single extra electron on any of the
two dots will be renormalized from its initial value of $\epsilon $,
since the bath relaxes to a ground state of lower energy in presence
of the electron. We will assume that the value of $\epsilon $ has
been chosen exactly to compensate for this energy change, which is
given by $-\int _{0}^{\infty }d\omega \, \left\langle \hat{F}\hat{F}\right\rangle _{\omega }/\omega $
(see App. \ref{indepBosonApp}). Then, the energy of an electron on
the dot (and the bath in its new ground state) is the same as that
of the electron being in the lead, at the Fermi energy of $\epsilon _{F}\equiv 0$
(for $V=0$). 

Tunneling of an electron from the dots to the lead will not change
the bath state, but it will displace the origin of the harmonic oscillators
comprising the bath, since the coupling to $\hat{F}$ is switched
off ($\hat{n}_{+}-\hat{n}_{-}$ changes to zero). Therefore, the original
ground state of the bath (in presence of the electron) will become
a superposition of excited states in the new bath potential (in absence
of the electron; see Fig. \ref{DDevpic}). On the other hand, since
energy conservation has to be fulfilled with respect to the total
energy of the electrons and the bath before and after the tunneling
event, only those excited bath states can be reached whose energies
are not greater than $eV$, the energy supplied to the electron by
the bias voltage. This leads to the Coulomb-blockade type \emph{suppression}
of the tunneling rate at low bias voltages, for $\varphi =0$. Physically,
this effect is just the same as that described by Franck-Condon overlap
integrals evaluated between vibronic states for electronic transitions
in molecules. Qualitatively, this effect is independent of the interference
setup, since it already occurs for tunneling through a single dot
coupled to a bath.

In contrast, for the case of destructive interference ($\varphi =\pi $),
the bath may actually \emph{enhance} the tunneling rate from its initial
value of $0$, since it partly destroys the phase coherence that is
a presupposition for perfect interference. An electron coming from
the left lead will form the following entangled state with the bath,
instead of the symmetric superposition $\left|e\right\rangle =(\left|+\right\rangle +\left|-\right\rangle )/\sqrt{2}$:

\begin{equation}
(\left|+\right\rangle \left|\chi _{+}\right\rangle +\left|-\right\rangle \left|\chi _{-}\right\rangle )/\sqrt{2}\, .\label{DDentangled}\end{equation}

Here the states $\left|\chi _{\pm }\right\rangle $ denote the respective
ground states of the bath for a bath Hamiltonian given by $\hat{H}_{B}\pm \hat{F}$,
which are related to each other by a parity transformation (This also
means we assume by definition there to be no phase factor between
these states; e.g. both may be assumed to have real-valued positive
wave functions). Actually, the entangled state considered here will
be formed only if the electron is given barely enough energy to enter
the double-dot at all (i.e. chemical potential of the left lead infinitesimally
larger than the renormalized level position). Otherwise, excited bath
states may be created even at this step. These complications will
be taken care of in the complete discussion of the sequential tunneling
current (Section \ref{DDseqTunnelingSection}). There, it will turn
out that the tunneling decay rate derived in the following, based
on our physically motivated ansatz (\ref{DDentangled}), is exactly
the rate that enters the full master equation. Thus, we proceed with
the ansatz (\ref{DDentangled}) for the initial entangled state, in
order to calculate the rate for such an electron to tunnel into the
right lead.

The bath measures (to some extent) which dot the electron resides
on, such that the reduced system density matrix (for the electron
on the two dots) becomes mixed and its off-diagonal elements get suppressed
by the overlap factor $\left\langle \chi _{+}|\chi _{-}\right\rangle $.
Put differently, the phase factor between the two dot states in the
wave function of the electron (initially equal to $+1$) becomes uncertain.
Therefore, there is a finite probability of 

\begin{equation}
P_{o}=(1-\left\langle \chi _{+}|\chi _{-}\right\rangle )/2\label{eq:Podd}\end{equation}

to find the electron in the antisymmetric (odd) state $\left|o\right\rangle \equiv (\left|+\right\rangle -\left|-\right\rangle )/\sqrt{2}$.
At $\varphi =\pi $, where tunneling decay of the symmetric superposition
$\kt{e}$ is blocked due to destructive interference, the state $\kt{o}$
is allowed to decay into the lead, at the maximal rate of $2\Gamma _{0}$.
In this way, the interference-induced blockade of electron tunneling
is lifted by dephasing. 

However, this simple picture is true only for large bias voltages,
when energy conservation permits any final state of the bath after
the tunneling event. If the maximum energy supplied to the electron
is limited, the suppression discussed above (for the case of $\varphi =0$)
will apply again. In particular, if the bias voltage is turned to
zero, energy conservation only allows the state $\left|\chi _{0}\right\rangle $
to be reached, which is the ground state of the bath in the absence
of any electrons on the dots. Then, the tunneling rate is exactly
zero again, despite the fact that the reduced density matrix of the
electron may be mixed to a strong extent. The reason is the following:
When the overlap of the entangled state (\ref{DDentangled}) with
the state $\left|\chi _{0}\right\rangle $ is taken, the two overlap
factors $\left\langle \chi _{0}|\chi _{+}\right\rangle $ and $\left\langle \chi _{0}|\chi _{-}\right\rangle $
turn out to be the same, \emph{if} the coupling of the bath to the
two dots is symmetric (i.e. of equal strength, only of opposite sign),
which we have assumed in writing down the Hamiltonian, Eq. (\ref{DDhamiltonian}).
Therefore, the electronic state resulting from the projection of (\ref{DDentangled})
onto $\kt{\chi _{0}}$ is equal to the symmetric combination, whose
decay is forbidden. Thus, the combination of energy conservation and
Pauli blocking prevents a finite tunneling rate at zero bias voltage,
in spite of the mixed state of the electron coupled to the bath. In
this limit the entanglement between electron and bath only leads to
renormalization effects (such as the change in tunneling rate), but
not to genuine dephasing. If the coupling were asymmetric, then destructive
interference could be lost even without dephasing (merely due to renormalization),
just as it would be the case for initially asymmetric bare tunnel
couplings. That is why the asymmetric case is uninteresting for our
purposes of distinguishing renormalization effects from real dephasing. 

However, whether we are indeed able to claim that dephasing actually
vanishes in the limit of low bias voltages will depend on the behaviour
of the tunneling rate as a function of $V$ and on the comparison
of the cases $\varphi =0$ and $\varphi =\pi $. Here, the bath spectrum,
and, above all, its low-frequency properties, enter. In order to be
able to discuss $\Gamma (V,\varphi )$ quantitatively, we will make
use of the concepts of the $P(E)$ theory of tunneling in a dissipative
environment.

\section{Decay rate and connection to $P(E)$ theory}

\label{sec:Decay-rate-and}The tunneling rate $\Gamma $ will be calculated
using the standard Fermi Golden Rule, i.e. lowest order perturbation
theory in the bare tunneling rate $\Gamma _{0}$, but taking into
account exactly the bath coupling. In deriving the formula for $\Gamma $,
it turns out to be useful to assume that the bath oscillators do \emph{not}
get shifted in the tunneling event (unlike the qualitative considerations
from above), but it is rather the bath states which get displaced
(in the opposite direction). Obviously, this amounts to the same,
as long as we are interested only in overlap integrals of different
bath states after the event. To that end, we introduce the displacement
operator $\exp (i\hat{\phi })$, which transforms the bath ground
state of $\hat{H}_{B}$ into that of $\hat{H}_{B}+\hat{F}$. Here
$\hat{\phi }$ is a suitable hermitian operator that is linear in
the bosonic variables of the bath. In fact, this amounts to performing
the canonical transformation of the independent boson model\cite{mahan},
see Appendix \ref{indepBosonApp}. In terms of the two dot states
$+$ and $-$, we have $\hat{F}_{+}=\hat{F}$ and $\hat{F}_{-}=-\hat{F}$,
as well as $\ph _{+}=\ph $ and $\ph _{-}=-\ph $. The transformation
eliminates the system-bath coupling from the Hamiltonian, but gives
rise to modified dot operators $\hat{d}_{\pm }'=e^{\pm i\ph }\hat{d}_{\pm }$
in the transformed tunnel Hamiltonian $\hat{V}_{R}'$ (see Eq. (\ref{IBdtransform})). 

We will assume the tunnel-coupling to be sufficiently weak, such that
we can use lowest-order perturbation theory to calculate the tunneling
decay rate:

\begin{equation}
\Gamma =2\pi \sum _{f}\left|\left\langle f|\hat{V}'_{R}|i\right\rangle \right|^{2}\delta (E_{f}-E_{i})\, ,\end{equation}

where the initial state $\left|i\right\rangle $ is given by the configuration
involving the electron residing in the symmetric superposition on
the dots, the unperturbed Fermi sea in the lead and the bath in its
ground state $\left|i_{B}\right\rangle $. The bath ground state has
become independent of the position of the electron, due to the above-mentioned
transformation. At finite temperatures, an additional thermal average
over the initial bath state and the initial state of the electrons
in the lead has to be performed. The energies and eigenstates refer
to the Hamiltonian without tunnel coupling. Applying the new tunneling
Hamiltonian $\hat{V}'_{R}$ to the initial state, we obtain the following
expression:

\begin{eqnarray}
\Gamma =\pi \sum _{k,f_{B}}\left|t_{k}\right|^{2}(1-f(\epsilon _{k}+eV))\times  &  & \nonumber \\
\left|\left\langle f_{B}|e^{+i\hat{\phi }}+e^{i\varphi }e^{-i\hat{\phi }}|i_{B}\right\rangle \right|^{2}\delta (E_{f}^{B}-E_{i}^{B}+\epsilon _{k})\, , &  & \label{DDgammaanfang}
\end{eqnarray}

Here $f(\cdot )$ is the Fermi function (for chemical potential equal
to zero), and $E_{f,i}^{B}$ are the energies of the initial and final
bath states. The energy supplied to the bath is equal to the energy
lost by the electron (given by $-\epsilon _{k}$, since the renormalized
dot energy is zero). Following the usual derivation of the $P(E)$
theory \cite{ingoldhabil,SchoenPE}, we express the energy-conserving
$\delta $ function as an integral over time and also replace the
sum over lead states $k$ by an integral over the energy $E=-\epsilon _{k}$
supplied to the bath, finally yielding:

\begin{eqnarray}
\Gamma =\Gamma _{0}\int _{-\infty }^{+\infty }dE(1-f(-E+eV))\int _{-\infty }^{+\infty }\frac{dt}{2\pi }e^{iEt}\times  &  & \nonumber \\
\frac{1}{2}\left\langle (e^{-i\hat{\phi }(t)}+e^{-i\varphi }e^{i\hat{\phi }(t)})(e^{i\hat{\phi }}+e^{i\varphi }e^{-i\hat{\phi }})\right\rangle  &  & \label{DDGammavorher}
\end{eqnarray}

For the case of arbitrary temperature, the brackets denote a thermal
average over the initial bath state $\left|i_{B}\right\rangle $.
We introduce the definitions:

\begin{equation}
P_{(-)}(E)=\frac{1}{2\pi }\int _{-\infty }^{+\infty }dt\, e^{iEt}\, e^{\pm \left\langle \hat{\phi }(t)\hat{\phi }\right\rangle -\left\langle \hat{\phi }^{2}\right\rangle }\, .\label{DDPEDef}\end{equation}

This permits us to write down our final result for the tunneling decay
rate in terms of $P_{(-)}(E)$:

\begin{equation}
\Gamma =\Gamma _{0}\int _{-\infty }^{+\infty }dE\, (1-f(-E+eV))\, (P(E)+\cos (\varphi )P_{-}(E))\label{DDGammabasic}\end{equation}

The formula given here constitutes the basic expression for the decay
rate as a function of bias voltage and interference phase $\varphi $.
It represents the appropriate modification of Eq. (\ref{eq:nobath})
in presence of a bath. 

Note that for the slightly more general case of arbitrarily correlated
fluctuating potentials $\hat{F}_{+}$ and $\hat{F}_{-}$ attached
to the dots (i.e. an interaction of the form $\hat{F}_{+}\hat{n}_{+}+\hat{F}_{-}\hat{n}_{-}$),
the function $P_{-}(E)$ would contain the cross-correlator of the
associated phases $\hat{\phi }_{+}$ and $\hat{\phi }_{-}$, while
$P(E)$ would depend on the autocorrelator of $\hat{\phi }_{+}$ or
$\hat{\phi }_{-}$ (assumed to be the same, for the setup to remain
symmetric). In contrast to the model treated here, such an interaction
would also involve fluctuations of the sum of energies of the dot-levels.
However, they would only add to the renormalization effects mentioned
previously and do not contribute to dephasing by themselves, since
such fluctuations cannot distinguish between the two interfering paths.

By using the definitions

\begin{equation}
\gamma _{(-)}\equiv \Gamma _{0}\int dE\, (1-f(-E+eV))\, P_{(-)}(E)\, ,\label{DDgammaDefinition}\end{equation}

we can write

\begin{equation}
\Gamma =\gamma +\cos (\varphi )\gamma _{-}\, .\end{equation}

The strength of the dependence of $\Gamma $ on the phase $\varphi $
may be taken as a signature of phase coherence in our model. We define
the {}``visibility'' of the interference pattern in the usual way,
by

\begin{equation}
\upsilon \equiv (\Gamma _{max}-\Gamma _{min})/(\Gamma _{max}+\Gamma _{min}),\label{eq:visDef}\end{equation}

which is equal to the ratio

\begin{equation}
\upsilon =\frac{\gamma _{-}}{\gamma }\, .\label{eq:visGamm}\end{equation}

The visibility $\upsilon $ will be $1$ whenever the destructive
interference is perfect, and it is zero if there is no dependence
of $\Gamma $ on $\varphi $.

The effects of the bath on the decay rate are encoded in the functions
$P(E)$ and $P_{-}(E)$, whose general properties we will discuss
now. In the next section, we will evaluate them for different types
of bath spectra. 

As usual, the function $P(E)$ describes the probability (density)
that an electron will emit the energy $E$ into the bath while tunneling
into the lead. It is real, nonnegative and normalized to unity \cite{SchoenPE,ingoldhabil}. 

At large times $\left|t\right|\rightarrow \infty $, the correlation
function $\left\langle \hat{\phi }(t)\hat{\phi }\right\rangle $ in
the exponent of the integral (\ref{DDPEDef}) will decay towards zero,
for a continuous bath spectrum. This means that the integrand of $P(E)$
approaches the value of $z\equiv \exp (-\left\langle \hat{\phi }^{2}\right\rangle )$,
starting from $1$ at $t=0$. Therefore, $P(E)$ contains a {}``quasiparticle
$\delta $ peak'' of strength $z$ at $E=0$, if $z$ does not vanish.
It corresponds to the probability $z$ of having no energy transfer
at all from the electron to the bath (similar to the recoil-free emission
of a $\gamma $ ray by a nucleus inside a crystal, i.e. the Mössbauer
effect).

The function $P_{-}(E)$ in front of the $\cos (\varphi )$ term in
Eq. (\ref{DDGammabasic}) is different: The integrand of $P_{-}(E)$
will increase at large times, towards the value of $z$, starting
from $z^{2}$ at $t=0$. The function $P_{-}(E)$ is real-valued (because
of $\left\langle \hat{\phi }(t)\hat{\phi }\right\rangle =\left\langle \hat{\phi }\hat{\phi }(t)\right\rangle ^{*}$),
but it can become negative. Therefore, it cannot be interpreted as
a probability density, in contrast to $P(E)$. Its normalization is
given by:

\begin{equation}
\int dE\, P_{-}(E)=z^{2}\, .\end{equation}

If $z$ is nonzero, $P_{-}(E)$ also has a $\delta $ peak at $E=0$,
of weight $z$, just as $P(E)$. As a consequence, in the case of
destructive interference ($\varphi =\pi $), the tunneling rate $\Gamma $
at $V\rightarrow 0,\, T=0$ still vanishes even in the presence of
the bath, since the $\delta $ peaks contained in $P(E)$ and $P_{-}(E)$
cancel exactly in the integral (\ref{DDGammabasic}). The physical
reason for this coherence has been discussed at the end of the previous
section. 

In the case of constructive interference ($\varphi =0$), at $T=0$
and for $V\rightarrow 0$, the integration over $E$ will only capture
the $\delta $ peaks contained in $P_{(-)}(E)$, yielding $\Gamma =2z\Gamma _{0}$.
Thus, the tunneling rate is suppressed by the constant factor $z$
from its noninteracting value. However, this may be interpreted as
a mere renormalization of the effective tunnel coupling, since the
visibility $\upsilon $ of the interference pattern is still equal
to unity. In order to connect this result to the qualitative discussion
from above, we note that the overlap of the two different bath ground
states that are adapted to the absence or presence of an electron
on dot $\pm $, is given by: 

\begin{equation}
\left\langle \chi _{0}|\chi _{\pm }\right\rangle =\left\langle \chi _{0}\left|e^{\pm i\hat{\phi }}\right|\chi _{0}\right\rangle =\exp (-\left\langle \hat{\phi }^{2}\right\rangle /2)=z^{1/2}\, ,\end{equation}

Therefore, the magnitude squared of this overlap, that determines
the probability of tunneling without exciting any bath mode, is equal
to $z$.

On the other hand, for sufficiently large bias voltages (much larger
than the cutoff frequency of the bath spectrum), the normalization
conditions for $P_{(-)}(E)$ yield

\begin{equation}
\Gamma =\Gamma _{0}(1+z^{2}\cos (\varphi ))\, .\label{DDgammahighv}\end{equation}

The visibility is given by $\upsilon =z^{2}$. In this limiting case,
where the restrictions due to energy conservation and the Pauli principle
are no longer important, the tunneling rate $\Gamma $ at the point
$\varphi =\pi $ of destructive interference does not vanish. It takes
the value $\Gamma _{0}(1-z^{2})$, which is small if the effects of
the bath are weak ($z$ near to $1$) and is equal to one half the
ideal maximum value $2\Gamma _{0}$ for a bath that is sufficiently
strong to destroy phase coherence completely ($z=0$), leading to
an incoherent mixture of symmetric and antisymmetric states on the
two dots. In the latter case, the visibility vanishes (even for arbitrary
voltages), since then $P_{-}(E)$ is equal to zero, which makes $\Gamma $
independent of $\varphi $. This will be true for the Ohmic bath,
to be discussed in the next section. 

As explained above, the reduced density matrix of the electron on
the dots coupled to the bath predicts a finite probability of $P_{o}=(1-\left\langle \chi _{+}|\chi _{-}\right\rangle )/2$
to find the electron in the antisymmetric state if one starts out
from the symmetric superposition before coupling it to the bath. The
overlap factor of the bath states involved in this probability can
be expressed as 

\begin{equation}
\left\langle \chi _{+}|\chi _{-}\right\rangle =\left\langle \chi _{0}|(e^{-i\hat{\phi }})^{2}|\chi _{0}\right\rangle =z^{2}\, .\label{DDchipm}\end{equation}

Comparing with the result $\Gamma (\varphi =\pi )=\Gamma _{0}(1-z^{2})$
given above, it may be observed that the decay rate at sufficiently
large bias voltages is indeed determined directly by the probability
to find the electron in the state whose decay is not forbidden by
destructive interference (as has been argued already at the end of
the previous section, near Eq. (\ref{eq:Podd})). It is only in this
limiting case, where an arbitrary amount of energy is available for
excitation of the bath, that the suppression of interference effects
in the transport situation is correctly deduced from the electron's
reduced density matrix in the presence of the bath. Formally, this
holds because the sum over final bath states $f_{B}$ in Eq. (\ref{DDgammaanfang})
is not restricted any more and corresponds to the insertion of a complete
set of basis states. Thus, one obtains, directly from Eq. (\ref{DDgammaanfang}):

\begin{equation}
\Gamma =\frac{\Gamma _{0}}{2}\left\langle \chi _{+}+e^{-i\varphi }\chi _{-}|\chi _{+}+e^{i\varphi }\chi _{-}\right\rangle \, ,\label{eq:highvexplain}\end{equation}

which reduces to Eq. (\ref{DDgammahighv}) when the overlaps are evaluated,
using Eq. (\ref{DDchipm}). Physically, the case of high bias voltage
corresponds to a kind of infinitely fast von Neumann projection measurement
that determines the state of the electron, revealing the fluctuations
due to the bath. In contrast, at low bias voltages (low energy supply),
a kind of {}``weak'' measurement is carried out that takes a longer
amount of time, such that only the low-frequency fluctuations of the
bath are important for dephasing.

\section{Evaluation for different bath spectra}

\label{sec:Evaluation-for-different}We will restrict the discussion
to $T=0$ at first. 

The simplest example for the bath is a single harmonic oscillator
of frequency $\omega $. This offers an approximate description of
the interaction with optical phonon modes ({}``Einstein model'').
In this case, $P(E)$ and $P_{-}(E)$ can be obtained easily by expanding
the exponential in a Taylor series and using $\left\langle \hat{\phi }(t)\hat{\phi }\right\rangle =\left\langle \hat{\phi }^{2}\right\rangle \exp (-i\omega t)$,
before the integration over time is performed. For $P(E)$, the resulting
series of $\delta $ peaks at harmonics of $\omega $ corresponds
to all possible processes where the electron emits any number $n$
of phonons into the bath while tunneling into the lead. The expression
for $P_{-}(E)$ is the same, apart from alternating signs in front
of the $\delta $ functions: 

\begin{equation}
P_{(-)}(E)=z\sum _{n=0}^{\infty }\frac{\left\langle \pm \hat{\phi }^{2}\right\rangle ^{n}}{n!}\delta (E-n\omega )\, .\end{equation}

Thus, every process involving the transfer of an even number of quanta
to the bath will not ruin the destructive interference at $\varphi =\pi $,
since the corresponding contributions from $P(E)$ and $P_{-}(E)$
cancel in Eq. (\ref{DDGammabasic}). This is because the coupling
between electron and bath is of the type $(\hat{n}_{+}-\hat{n}_{-})\hat{F}$,
which gives a different sign of the interaction amplitude for a phonon
emission process, depending on the dot. Therefore, the amplitude of
emission of an \emph{even} number of phonons will \emph{not} depend
on the dot, it is insensitive to the state of the electron, and the
amplitudes of the electron tunneling from $\left|+\right\rangle $
and $\left|-\right\rangle $ will still interfere destructively.

In contrast, emission processes involving an odd number of quanta
introduce a negative sign for an electron starting in state $\left|-\right\rangle $,
{}``detecting'' the path (or rather, the initial state) of the electron
and interfering \emph{constructively} with the processes from $\left|+\right\rangle $.
This lifts the destructive interference and makes $\Gamma \neq 0$
at $\varphi =\pi $. However, below the frequency $\omega $ of the
oscillator, destructive interference at $\varphi =\pi $ is still
perfect since no quantum can be emitted, while the magnitude of $\Gamma $
at $\varphi =0$ is renormalized by the factor $z$, as has been discussed
above in general for the limiting case $V\rightarrow 0$. The same
holds true for any bath with a finite excitation gap, at $T=0$. This
is shown in Figs. \ref{DDGamma0} and \ref{DDGammaPi}, to be discussed
in the next section.

We now pass on to arbitrary bath spectra. At first, we will cover
the case $z\neq 0$ ({}``weak baths''), when we can apply perturbation
theory to discuss the behaviour of $P_{(-)}(E)$ at low energy transfers
$E$ (and, consequently, that of $\Gamma $ at low voltages). A Taylor-expansion
of the exponent in Eq. (\ref{DDPEDef}) yields:

\begin{eqnarray}
P_{(-)}(E)=\frac{z}{2\pi }\sum _{n=0}^{\infty }\frac{1}{n!}\int _{-\infty }^{+\infty }dt\, e^{iEt}\left[\pm \left\langle \hat{\phi }(t)\hat{\phi }\right\rangle \right]^{n} &  & \nonumber \\
=z\sum _{n=0}^{\infty }\frac{(\pm 1)^{n}}{n!}(\left\langle \hat{\phi }\hat{\phi }\right\rangle _{\omega }*\ldots *\left\langle \hat{\phi }\hat{\phi }\right\rangle _{\omega })(E) &  & \label{DDPfolding}
\end{eqnarray}

The repeated convolution product contains $n$ times the correlator
$\left\langle \hat{\phi }\hat{\phi }\right\rangle _{\omega }$, for
$n=0$ it is to equal $\delta (E)$, and the negative sign holds for
$P_{-}(E)$. 

For the following discussion, we prescribe the spectrum of the fluctuating
potential $\hat{F}$ to be a power-law in frequency $\omega $ (at
$T=0$), with exponent $s$:

\begin{equation}
\left\langle \hat{F}\hat{F}\right\rangle _{\omega }^{T=0}=2\alpha \omega _{c}\left(\frac{\omega }{\omega _{c}}\right)^{s}\theta (\omega _{c}-\omega )\theta (\omega ),\label{DDUUlaw}\end{equation}

The dimensionless parameter $\alpha $ characterizes the bath strength.
In order to be able to rely on perturbation theory, we have to ensure
$z>0$. Since $\left\langle \hat{\phi }\hat{\phi }\right\rangle _{\omega }=\left\langle \hat{F}\hat{F}\right\rangle _{\omega }/\omega ^{2}$,
the variance of the fluctuating phase, $\left\langle \hat{\phi }^{2}\right\rangle $,
will be finite only for $s>1$ (at $T=0$, otherwise $s>2$). In that
case, we have $z=\exp (-2\alpha /(s-1))$. This means the perturbative
analysis presented above is restricted to a super-Ohmic bath, $s>1$.
The case of the Ohmic bath will be discussed separately further below.

After keeping only terms up to second order in the expansion of $P_{(-)}(E)$
given in Eq. (\ref{DDPfolding}), we get

\begin{equation}
P(E)+P_{-}(E)=z(2\delta (E)+(\left\langle \hat{\phi }\hat{\phi }\right\rangle _{\omega }*\left\langle \hat{\phi }\hat{\phi }\right\rangle _{\omega })(E)+\ldots )\, ,\end{equation}

for the symmetric combination, that will determine the prefactor of
$1+\cos (\varphi )$ in the expression for $\Gamma $, Eq. (\ref{DDGammabasic}),
and

\begin{equation}
P(E)-P_{-}(E)=2z\left\langle \hat{\phi }\hat{\phi }\right\rangle _{E}+\ldots \end{equation}

for the antisymmetric combination (determining the prefactor of $1-\cos (\varphi )$).
Inserting these into (\ref{DDGammabasic}), using the power law for
$\left\langle \hat{\phi }\hat{\phi }\right\rangle _{\omega }=\left\langle \hat{F}\hat{F}\right\rangle _{\omega }/\omega ^{2}$
given by (\ref{DDUUlaw}), and performing the energy integrals, we
find, for sufficiently low voltages ($2\alpha (eV/\omega _{c})^{s-1}\ll s-1$): 

\begin{eqnarray}
\Gamma \approx \frac{\Gamma _{0}}{2}z\{(1+\cos (\varphi ))(1+\frac{\alpha ^{2}C_{s}}{(s-1)}\left(\frac{eV}{\omega _{c}}\right)^{2(s-1)})+ &  & \nonumber \\
(1-\cos (\varphi ))\frac{2\alpha }{s-1}\left(\frac{eV}{\omega _{c}}\right)^{s-1}\}\, . &  & \label{DDweakgamma}
\end{eqnarray}

The numerical prefactor $C_{s}$ is defined as $\int _{0}^{1}(y(1-y))^{s-2}dy$. 

From Eq. (\ref{DDweakgamma}), we see that the destructive interference
at $\varphi =\pi $ is perfect at $V=0$, but gets lifted when increasing
the bias voltage, with a power $V^{s-1}$. In contrast, the decay
rate $\Gamma $ at $\varphi =0$ starts out from the constant value
of $2z\Gamma _{0}$ and grows as $V^{2(s-1)}$. Therefore, the visibility
$\upsilon $ starts out at $1$ for $V=0$ but decreases as:

\begin{equation}
\upsilon \approx 1-\frac{4\alpha }{s-1}\left(\frac{eV}{\omega _{c}}\right)^{s-1}\, .\label{viasapprox}\end{equation}

For $s\downarrow 1$, the range in bias voltage $V$ where these approximate
expressions hold shrinks to zero (at constant $\alpha $ and $\omega _{c}$).
At $s=1$, i.e. for the Ohmic bath, the probability $z$ of not emitting
energy into the bath vanishes completely. As discussed above, this
means that there is no $\varphi $-dependence at all in $\Gamma $,
and, consequently, the visibility is zero at all bias voltages. Furthermore,
the tunneling rate vanishes for $eV\rightarrow 0$, even at $\varphi =0$.
This is the well-known Coulomb-blockade type of behaviour for tunneling
in presence of Ohmic dissipation \cite{devoret}. At higher bias voltages,
the blockade is removed and $\Gamma $ grows towards $\Gamma _{0}$.
The growth at low voltages is determined by the power-law behaviour
of $P(E)$, which rises as $c\omega _{c}^{-2\alpha }E^{2\alpha -1}$,
where the exponent is determined by the bath-strength rather than
the exponent $s=1$ of the bath spectrum. The dimensionless prefactor
$c$ must be found from the normalization condition for $P(E)$ and
depends only on $\alpha $ (and the type of cutoff in the bath spectrum).
Therefore, in the case of the Ohmic bath we have, at low $V$ and
$T=0$: 

\begin{equation}
\Gamma (V)=\Gamma _{0}\frac{c}{2\alpha }\left(\frac{eV}{\omega _{c}}\right)^{2\alpha }\, .\end{equation}

Finally, we briefly discuss the case of finite temperatures, $T>0$.

In that case, the variance of $\hat{\phi }$ is given by

\begin{equation}
\left\langle \hat{\phi }^{2}\right\rangle =\int _{0}^{\infty }d\omega \, \left\langle \hat{\phi }\hat{\phi }\right\rangle _{\omega }^{(T=0)}\coth \left(\frac{\omega }{2T}\right)\, ,\end{equation}

which yields

\begin{equation}
\left\langle \hat{\phi }^{2}\right\rangle \approx \left\langle \hat{\phi }^{2}\right\rangle ^{(T=0)}+4\alpha \left(\frac{T}{\omega _{c}}\right)^{s-1}\int _{0}^{\infty }\frac{y^{s-2}}{e^{y}-1}dy\, .\end{equation}

The approximation of extending the integral to infinity holds for
temperatures much smaller than the bath cutoff $\omega _{c}$. This
formula gives the temperature-dependence of the renormalization factor
$z=\exp \left(-\left\langle \hat{\phi }^{2}\right\rangle \right)$.
The second integral diverges for $s\leq 2$, because $z=0$ for these
cases, in contrast to $T=0$ where $z=0$ only for $s\leq 1$. Again,
this results in complete absence of the interference effect in the
tunneling rate $\Gamma (V,\varphi )$ (because $P_{-}(E)$ vanishes).
It may seem surprising that an infinitesimally small temperature can
yield such a drastic qualitative change (for $1<s\leq 2$), compared
to the zero-temperature case, since the difference should be observable
only at very large times $t\gg 1/T$. However, it must be remembered
that our analysis is carried out for the limit $\Gamma _{0}\rightarrow 0$,
where the average decay time of the given state is inifinitely large.
In other words, the limits $T\rightarrow 0$ and $\Gamma _{0}\rightarrow 0$
do not commute for such relatively strong baths. At finite $\Gamma _{0}$,
the transition from one to the other case should turn out to be smooth,
but this goes beyond the present analysis. 

Apart from the change in $z$ with temperature, there are two other
important differences to the case $T=0$: First of all, even at $V\rightarrow 0$
the electron may emit energy into the bath, due to the thermal smearing
of the Fermi surface in the lead (lifting of Pauli blocking). Secondly,
it may now also absorb some energy during the tunneling process. Both
facts will, in general, lead to a finite tunneling decay rate at $\varphi =\pi ,\, V\rightarrow 0$
for any bath, where, at $T=0$, the rate had vanished in any case. 

We can approximate the visibility $\upsilon $ at $V\rightarrow 0$
and finite $T$ by using the expansion (\ref{DDPfolding}). Inserting
the resulting expressions for $\gamma _{(-)}$ (\ref{DDgammaDefinition})
into $\upsilon =\gamma _{-}/\gamma $, we obtain

\begin{equation}
\upsilon (T,V\rightarrow 0)\approx 1-4\int d\epsilon \, \left\langle \ph \ph \right\rangle _{\epsilon }f(\epsilon )\, .\end{equation}

We evaluate the integral for a power-law bath spectrum in the limit
$T\ll \omega _{c}$:

\begin{eqnarray}
\int d\epsilon \, \left\langle \ph \ph \right\rangle _{\epsilon }f(\epsilon ) & = & \nonumber \\
=\int _{0}^{\infty }d\epsilon \, \frac{\left\langle \ph \ph \right\rangle _{\epsilon }^{T=0}}{\sinh (\beta \epsilon )} &  & \nonumber \\
\approx 2\alpha \omega _{c}^{1-s}\int _{0}^{\infty }d\epsilon \, \frac{\epsilon ^{s-2}}{\sinh (\beta \epsilon )}\, . &  & 
\end{eqnarray}

This yields:

\begin{equation}
1-\upsilon (T,V\rightarrow 0)\approx 32\, \alpha \, \left(\frac{T}{\omega _{c}}\right)^{s-1}(\frac{1}{2}-2^{-s})\Gamma (s-1)\zeta (s-1)\, ,\label{DDvisfiniteT}\end{equation}

where $\Gamma $ is the Euler gamma function, and $\zeta $ the Riemann
zeta function. Therefore, the decrease of the visibility with increasing
temperature $T$ (and $V\rightarrow 0$) is governed by the same power-law
as that for increasing bias voltage $V$ at $T=0$, see Eq. (\ref{viasapprox}).

\section{Discussion of the results}

\label{sec:Discussion-of-the}The following discussion relates to
the results obtained for $T=0$, that are plotted in the figures.

\begin{figure}
\begin{center}\includegraphics[  height=7cm]{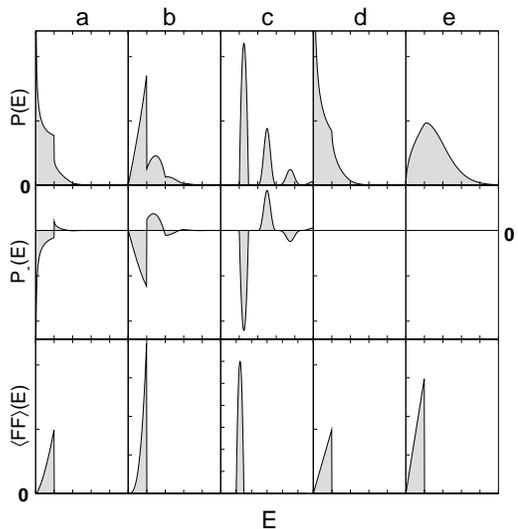}\end{center}

\caption{\label{DDbathspectra}The bath spectrum $\left\langle \hat{F}\hat{F}\right\rangle _{E}$
(bottom) and the resulting functions $P(E)$ (top) and $P_{-}(E)$
(middle), plotted vs. energy $E$, for different baths. Energies are
measured in units of the {}``bath cutoff'' $\omega _{c}$. Energy
axis is the same in all panels (starting at $E=0$, horizontal tick
distance: $1$); vertical tick distance in all panels is $0.5$. \emph{a}:
$s=1.5,\, \alpha =0.25$; \emph{b}: {}``acoustic phonons'', $s=3,\, \alpha =1$;
\emph{c}: {}``optical phonons'', Bath with gap; \emph{d}: $s=1,\, \alpha =0.25$;
\emph{e}: $s=1,\, \alpha =0.75$ (d,e are {}``Ohmic'' baths of different
strength, $z=0$)}
\end{figure}

\begin{figure}
\begin{center}\includegraphics[  height=7cm]{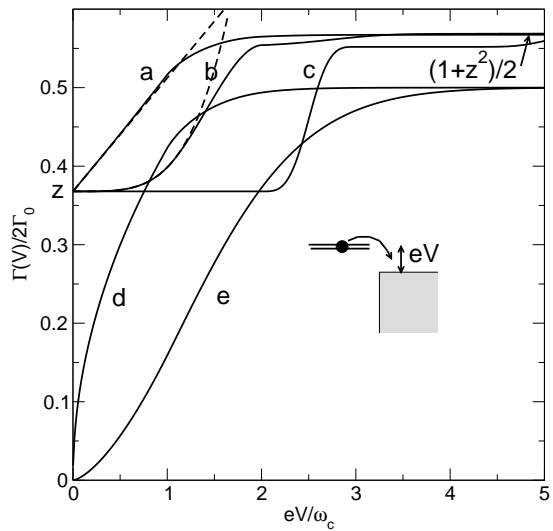}\end{center}

\caption{\label{DDGamma0}Decay rate $\Gamma $ as a function of bias voltage
$V$ for the case of \emph{constructive} interference ($\varphi =0$),
at $T=0$. Curves correspond to different bath spectra shown in Fig.
\ref{DDbathspectra}. Dashed lines correspond to approximation Eq.
(\ref{DDweakgamma}). The initial Coulomb-blockade type suppression
to a value of $\Gamma /2\Gamma _{0}=z$ ($z=0$ for the Ohmic bath
d,e) is lifted with increasing bias voltage, saturating at $\Gamma /2\Gamma _{0}=(1+z^{2})/2$.
Inset depicts energy diagram with definition of bias voltage for this
situation.}
\end{figure}

\begin{figure}
\begin{center}\includegraphics[  height=7cm]{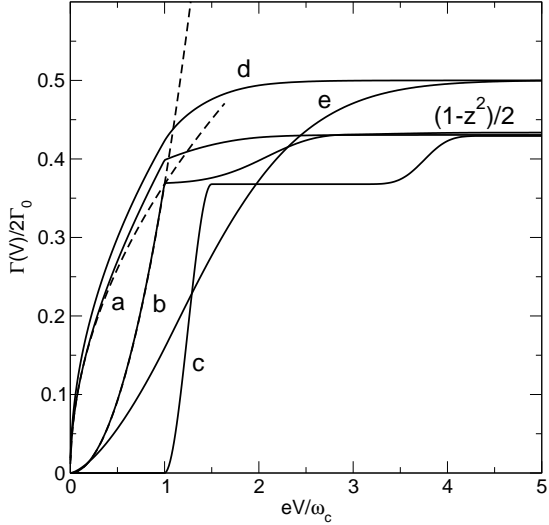}\end{center}

\caption{\label{DDGammaPi}Decay rate $\Gamma (V)$ for the case of \emph{destructive}
interference $(\varphi =\pi )$, at $T=0$. Dashed lines refer to
Eq. (\ref{DDweakgamma}). Due to dephasing, the decay rate becomes
finite at finite voltages, saturating at $\Gamma /2\Gamma _{0}=(1-z^{2})/2$.
For the Ohmic bath (d,e) the dependence is exactly equal to that for
$\varphi =0$ (Fig. \ref{DDGamma0}).}
\end{figure}
\begin{figure}
\begin{center}\includegraphics[  height=7cm]{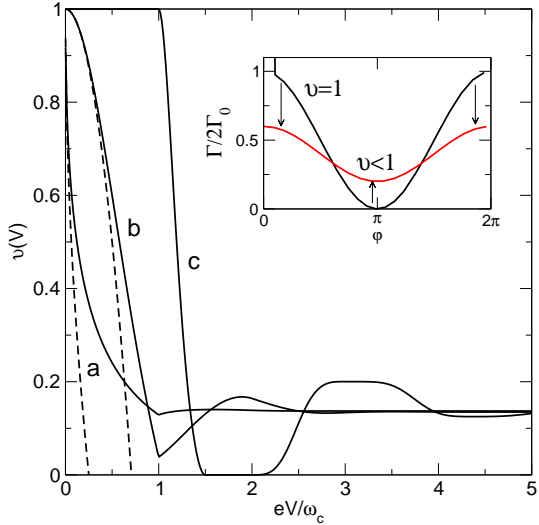}\end{center}

\caption{\label{DDVis}Visibility $\upsilon =(\Gamma _{\varphi =0}-\Gamma _{\varphi =\pi })/(\Gamma _{\varphi =0}+\Gamma _{\varphi =\pi })$
as a function of bias voltage $V$ for different bath spectra (see
Fig. \ref{DDbathspectra}). For the Ohmic bath (cases d,e) $\upsilon \equiv 0$.
Dashed lines correspond to Eq. (\ref{viasapprox}). Inset illustrates
change in interference pattern $\Gamma (\varphi )$ upon switching
on the interaction with the bath.}
\end{figure}

In Fig. \ref{DDbathspectra}, several different types of bath spectra
$\left\langle \hat{F}\hat{F}\right\rangle _{E}$ are shown. Cases
(a),(b),(d) and (e) are power-laws of the form given in Eq. (\ref{DDUUlaw}),
for a cutoff frequency of $\omega _{c}=1$. The last two (d,e) are
of Ohmic type ($s=1,\, z=0$), which corresponds physically to gate
voltage fluctuations due to Nyquist noise. Case (c) represents a bath
with an excitation gap (for example optical phonons), with a spectrum
given by an inverted parabola. In the limit of infinitely small spectral
bandwidth, it would correspond to the single harmonic oscillator (Einstein
mode) discussed above. Case (b), with a bath spectrum rising as $\omega ^{3}$,
corresponds to the experimentally relevant case of piezoelectric coupling
to acoustic phonons, which was determined to be the major inelastic
mechanism in the experiments of Ref.~\onlinecite{KouwenhouwenDD} on double-dots
in GaAs (see Ref.~\onlinecite{BrandesKramer} for a theoretical analysis
deriving this spectrum for wavelengths larger than the dot distance).
The spectra for the first three cases (a,b,c) have been chosen to
give the same renormalization factor, $z=1/e$. The same figure shows
the resulting functions $P(E)$ and $P_{-}(E)$. These have been obtained
using the integral equation described in Refs.~\onlinecite{minnhagen,ingoldhabil}.
We recall that the low-energy behaviour of $P(E)$ is given by $\left\langle \hat{\phi }\hat{\phi }\right\rangle _{E}=\left\langle \hat{F}\hat{F}\right\rangle _{E}/E^{2}$
for the cases with $z\neq 0$, where perturbation theory may be applied.
In case (c), the alternating signs of the different contributions
to $P_{-}(E)$ may be observed, whose physical meaning has been explained
above for the limiting case of the harmonic oscillator.

We now briefly mention some numerical estimates for the bath strengths
as they may occur in experimental situations.

In GaAs, the lack of inversion symmetry leads to piezoelectric fields
proportional to the lattice deformation, whose effect on electrons
at low frequencies is much stronger than that of the usual deformation
potential (where it is only the \emph{potential} that is proportional
to the deformation). For the piezoelectric coupling\cite{gant} to
acoustic phonons in GaAs, one finds (compare Ref.~\onlinecite{BrandesKramer})
$\left\langle \hat{\phi }\hat{\phi }\right\rangle _{\omega }^{T=0}=W\, \omega /(c_{s}/d)^{2}$
for $\omega \ll c_{s}/d$, where $c_{s}\approx 5\cdot 10^{3}\, m/s$
is an estimate for the average velocity of longitudinal sound waves
in GaAs, and $d$ denotes the distance between the quantum dots. We
obtain $W=const\cdot (eh_{14}/4\pi )^{2}/(\hbar \rho c_{s}^{3})$,
where $eh_{14}=1.4\, eV/nm$ is the single piezo-electric modulus
in the cubic $T_{d}$ structure of $GaAs$ and $\rho =5.3\cdot 10^{3}\, kg/m^{3}$
the mass density. The numerical constant is of order $1$ and accounts
for the details of the sound wave dispersion relation as well as the
orientation of the crystal axes with respect to the vector separating
the quantum dots. Inserting these values, $W$ is found to be on the
order of $0.01$. In order to obtain the renormalization factor $z$,
the spectrum $\left\langle \hat{\phi }\hat{\phi }\right\rangle _{\omega }$
must be integrated over all frequencies (see above), i.e. up to the
cutoff frequency $\omega _{c}$. The effective cutoff frequency $\omega _{c}\propto c_{s}/d_{0}$
is determined by the extent $d_{0}$ of the dot wave functions (for
$d_{0}=100nm$ one obtains $\omega _{c}\sim 50\, GHz$). Given the
present values, and assuming $d_{0}\approx d$, this leads to estimates
for $\int \left\langle \hat{\phi }\hat{\phi }\right\rangle _{\omega }d\omega $
on the order of $0.01$, yielding $z=\exp (-\left\langle \hat{\phi }^{2}\right\rangle )$
near $1$. Note that the distance $d$ between the dots cancels in
the estimate for $z$, as long as the cutoff frequency is assumed
to be given by $\omega _{c}\propto c_{s}/d$. However, as $\omega _{c}$
might be considerably larger than $c_{s}/d$ (if $d_{0}\ll d$), one
could also obtain a $z$ that deviates more strongly from unity.

For the Ohmic bath, we may imagine the quantum dots placed inside
a capacitor $C$ connected to a circuit of resistance $R$, such that
the potential difference $2\hat{F}$ between the dots would be given
by the fluctuating voltage drop across the capacitor. This leads to
a bath spectrum $\left\langle \hat{F}\hat{F}\right\rangle _{\omega }^{T=0}=\pi (R/R_{Q})\hbar ^{2}\omega /(1+(RC\omega )^{2})$,
with $R_{Q}=h/e^{2}$ the quantum of resistance. Therefore, the dimensionless
coupling constant $\alpha $ introduced above would be equal to $\alpha =(\pi /2)R/R_{Q}$,
which can have values both larger and smaller than $1$. 

Finally, for optical phonons, we use the Fröhlich interaction Hamiltonian
(Ref. \onlinecite{mahan}) with a dimensionless Fröhlich coupling
constant of $\alpha =0.07$ (GaAs) to obtain the rough estimate $\left\langle \hat{F}\hat{F}\right\rangle _{\omega }^{T=0}=\delta (\omega -\omega _{LO})\cdot (1meV)^{2}\cdot (100nm/d_{0})$,
with $\omega _{LO}\approx 5\cdot 10^{13}Hz$. This yields a $z$ deviating
from unity by about $10^{-3}$.

However, in the plots we have chosen $z=1/e$ for illustrative purposes.

The resulting behaviour of $\Gamma (\varphi ,V)$ at $T=0$, calculated
from Eq. (\ref{DDGammabasic}), is shown in Figs. \ref{DDGamma0}
and \ref{DDGammaPi}. In the case of constructive interference ($\varphi =0$,
Fig. \ref{DDGamma0}), the decay rate for the {}``weak baths'' (a,b,c)
starts out from $\Gamma /2\Gamma _{0}=z$ at $V=0$ and goes to $\Gamma /2\Gamma _{0}=(1+z^{2})/2$
at $eV/\omega _{c}\gg 1$. The initial deviation from the constant
value of $z$ at low voltages is given by the power-law $V^{2(s-1)}$
contained in Eq. (\ref{DDweakgamma}). In contrast, the decay rate
for the Ohmic bath (d,e) starts at $\Gamma =0$, rising with a power-law
and saturating at a value of $\Gamma /2\Gamma _{0}=1/2$, corresponding
to an equal admixture of odd and even states in the reduced density
matrix of the electron coupled to the bath. For destructive interference
($\varphi =\pi $, Fig. \ref{DDGammaPi}), the behaviour of (a) and
(b) at low voltages is given by $V^{s-1}$ (see Eq. (\ref{DDweakgamma})),
while the decay rate of the Ohmic bath (d,e) remains the same as that
for $\varphi =0$. In the special case (c) of the gapped bath, we
observe perfect destructive interference up to the excitation threshold
of the bath at $eV=\omega _{c}$, where $\Gamma (\varphi =\pi ,V)$
increases in a stepwise manner for the first time, with the next increase
at $eV=3\omega _{c}$. Note that, on the other hand, $\Gamma (\varphi =0,\, V)$
increases at even multiples of the excitation gap. The difference
comes about because it is only the emission of an odd number of phonons
into the bath that reveals the location of the electron, as discussed
above. This feature would be absent if the two dots were coupled to
two independent baths, whereas the other qualitative properties would
remain the same.

From the decay rates at $\varphi =0$ and $\varphi =\pi $, we may
calculate the visibility $\upsilon $ of the {}``interference pattern''
that is defined by the dependence of $\Gamma $ on $\varphi $. The
result is shown in Fig. \ref{DDVis}. As we have noted before, the
visibility is always zero for the Ohmic bath. On the other hand, for
the {}``weak baths'', it is perfect (equal to $1$) at $V\rightarrow 0$,
due to the perfect destructive interference, regardless of the suppression
factor $z$ appearing in $\Gamma (\varphi =0)$. In general, the visibility
decreases towards higher bias voltages before saturating at the limiting
value of $z^{2}$. However, in contrast to intuitive expectation,
the decrease may be nonmonotonous, i.e. the visibility of the interference
effect may actually be enhanced by increasing the supply of energy
available to the electron, although the decay rate $\Gamma $ always
increases monotonously at any $V$. This is particularly striking
in case (c), where the visibility drops down to zero in a certain
range before rising again. The decrease down to the exact value of
$0$ is related to the special choice of $\left\langle \hat{\phi }^{2}\right\rangle =1$
($z=1/e$), which gives equal strengths of the peak at $E=0$ and
the first peak around $E=\omega _{c}$, which then are able to cancel
in the integral $\gamma _{-}$ over $P_{-}(E)$ that is proportional
to the visibility. However, the physical reason for a dip in visibility
is rather generic: In that energy range, the decay rate $\Gamma $
for $\varphi =\pi $ has already increased due to dephasing, while
the blockade-type suppression of the value of $\Gamma $ for $\varphi =0$
has not yet been lifted. This is a consequence of the even-odd effect
discussed above.

\section{Sequential tunneling through the double-dot}

\label{DDseqTunnelingSection}Up to now, we have discussed in detail
the influence of the bath on the tunneling decay rate of an electron
which has been placed onto the two dots in the symmetric superposition.
In order to complete the picture, we have to calculate the sequential
tunneling current through such a double-dot interference setup. This
will be done by deriving and solving a master equation for the reduced
density matrix of the double-dot system, taking into account the system-bath
coupling exactly, while the tunnel-coupling is treated in leading
order. We are interested specifically in the nonlinear response, i.e.
in how an increasing bias voltage helps to destroy the phase coherence.
The tunneling rates calculated previously will serve as input to the
master equation.

However, in order to facilitate the understanding of the results,
we first turn to a qualitative description of the situation without
the bath.

At $\varphi =\pi $, tunneling is completely blocked, since the left
reservoir only couples to the even state $\kt{e}$, while the right
reservoir couples to the antisymmetric (odd) superposition, $\kt{o}$.
At $\varphi =0$, both reservoirs couple to $\kt{e}$, whereas $\kt{o}$
is completely decoupled from the leads (compare the discussion in
Ref.~\onlinecite{BoeseHofstetterSchoeller}). This means that a current
may flow if $\kt{o}$ is empty. However, if $\kt{o}$ is filled, the
current vanishes, because double-occupancy is forbidden in our model.
Since there is no way to change the occupation of $\kt{o}$, the stationary
density-matrix of the double-dot at $\varphi =0$ will be any convex
combination of these two possibilities (at $T=0$, in the absence
of other relaxation paths). At any value of $\varphi $ in between
these extremes, there is always the state $\kt{\Psi }=(\kt{+}-e^{-i\varphi }\kt{-})/\sqrt{2}$,
whose decay into the right lead is blocked by destructive interference.
As there is a nonvanishing overlap between $\kt{\Psi }$ and the state
$\kt{e}$ which is reached by tunneling from the left lead, one will
observe an accumulation of population in $\kt{\Psi }$, until the
current is blocked again. This argument holds at $T=0$, while at
finite temperatures the electron can decay towards the left lead and
make a new attempt. Therefore, in this simple picture, the stationary
current at $T=0$ would be zero at any $\varphi $ except for $\varphi =0$,
where it is undefined. 

However, one has to take into account that the coupling to the reservoirs
does not only lead to decay but also to an effective tunnel coupling
between $\kt{+}$ and $\kt{-}$. Although this cannot change the blockade
of the current at $\varphi =\pi $ (leading only to an energy shift
of $\kt{e}$ vs. $\kt{o}$), it does lift the blockade at other values
of $\varphi $. This is because the blocked state $\kt{\Psi }$ is
no longer stationary, such that an electron will not remain there
forever. The degeneracy at $\varphi =0$ still remains. Therefore,
in the ideal case without coupling to a bath, we expect the current
to vanish at $\varphi =\pi $ and to rise towards a maximal amplitude
near $\varphi =0$. According to the previous argument, at $T=0$
this maximal amplitude will be determined by the effective tunnel-coupling
between the dot states. 

Introducing the bath will then lead to renormalization effects and
spoil the perfect destructive interference at higher values of the
bias voltage (or temperature), qualitatively in the same way as it
has been explained above. We will show that the actual visibility
$\upsilon _{I}$ of the current interference pattern $I(\varphi )$
is given by a monotonous function of the visibility $\upsilon $ introduced
above for the tunneling rate (at symmetric bias). 

We start with the Hamiltonian that is obtained after applying the
unitary transformation of the independent boson model (\ref{IBHprime})
onto our Hamiltonian (\ref{DDhamiltonian}):

\begin{equation}
\hat{H}'=\epsilon '(\hat{n}_{+}+\hat{n}_{-})+U'\hat{n}_{+}\hat{n}_{-}+\hat{H}_{B}+\hat{H}_{L}+\hat{H}_{R}+\hat{V}'\, \end{equation}

Here $\epsilon '$ is the (renormalized) energy of the two states,
which we will take to be $\epsilon '=0$ from now on. $U'$ is the
interaction constant that involves both the Coulomb repulsion as well
as the effective attractive interaction induced by the bath. We assume
$U,U'\gg T,eV$, such that double-occupancy is forbidden. 

The term which we will treat as a perturbation is given by $\hat{V}'$,
describing the tunneling to the left and the right leads in the presence
of the bath. It is the transformed version of $\hat{V}$ (compare
Eqs. (\ref{DDtunnelHam}) and (\ref{eq:DDtunnelL}) and Appendix \ref{indepBosonApp}),
where the additional fluctuating phase factors $\exp (\pm i\ph )$
have been introduced:

\begin{equation}
\hat{V}'=\sum _{j=l,r}\sum _{\alpha =+,-}\hat{j}_{\alpha }\hat{d}_{\alpha }+h.c.\, ,\label{DDseqtunnVprime}\end{equation}

where

\begin{eqnarray}
\hat{l}_{\pm } & = & e^{\pm i\ph }\hat{l}\\
\hat{l} & = & \sum _{k}t_{k}^{L}\hat{a}_{Lk}^{\dagger }\\
\hat{r}_{+} & = & e^{+i\ph }\hat{r}\\
\hat{r}_{-} & = & e^{-i\ph }e^{i\varphi }\hat{r}\\
\hat{r} & = & \sum _{k}t_{k}^{R}\hat{a}_{Rk}^{\dagger }\, .\label{DDseqrDef}
\end{eqnarray}

As usual, the current through the device does not only depend on the
rates for electrons to tunnel into and out of the dots, but also on
the stationary state which the system assumes in the nonequilibrium
situation, i.e. under an applied bias voltage.

We will now derive a master equation for the reduced density matrix
$\hat{\rho }$ of the double-dot system, which contains the populations
$\rho _{++},\rho _{--},\rho _{00}$ ({}``$0$'' denoting {}``no
electron'') and the coherences $\rho _{+-}$ and $\rho _{-+}$ (with
$\rho _{00}=1-\rho _{++}-\rho _{--}$, $\rho _{\alpha 0}=\rho _{0\alpha }=0$
for $\alpha \neq 0$, and $\rho _{-+}=\rho _{+-}^{*}$). We cannot
simply use the standard kind of master equation, since we have to
deal with two degenerate levels $\kt{+}$ and $\kt{-}$, and it is
important that a tunneling event may create a coherent superposition
of $\kt{+}$ and $\kt{-}$ (for example the even state $\kt{e}$).
Such a master equation - for degenerate levels - has also been employed
in Ref.~\onlinecite{koeniggefen} (without coupling to the bath, and evaluated
in the linear-response regime). The equation is different from that
employed in the {}``orthodox'' theory of sequential tunneling, where
no coherent superpositions are involved. Note that for a finite tunnel-coupling
the levels could be treated as degenerate as long as their energetic
distance is much smaller than the level-broadening due to tunneling.
However, as we consider the limit $\Gamma _{0}\rightarrow 0$, we
need to have exactly equal energies. Otherwise, the energy of the
hole that is created in the left electrode would betray the dot state
which the electron has entered, thus preventing any coherent superposition
to form.

Given the initial reduced density matrix $\hat{\rho }(0)$, and assuming
the state of the environment (bath and reservoirs) to be independent
of the electronic state on the dot at $t=0$, we obtain the time-evolution
$\hat{\rho }(t)$ by tracing over the environmental degrees of freedom
({}``$E$''):

\begin{eqnarray}
\hat{\rho }(t) & = & tr_{E}[\hat{T}e^{-i\int _{0}^{t}ds\, \hat{V}'(s)}\hat{\rho }(0)\otimes \hat{\rho }_{E}\tilde{\hat{T}}e^{i\int _{0}^{t}ds\, \hat{V}'(s)}]\nonumber \\
 & = & \hat{\rho }(0)-\nonumber \\
 &  & \int _{0}^{t}dt_{1}\int _{0}^{t_{1}}dt_{2}\, tr_{E}[\hat{V}'(t_{1})\hat{V}'(t_{2})\hat{\rho }(0)\otimes \hat{\rho }_{E}+h.c.]\nonumber \\
 &  & +\int _{0}^{t}dt_{1}\int _{0}^{t}dt_{2}\, tr_{E}[\hat{V}'(t_{1})\hat{\rho }(0)\otimes \hat{\rho }_{E}\hat{V}'(t_{2})]\, \, +\ldots 
\end{eqnarray}

Physically, by using factorized initial conditions, we neglect correlations
between subsequent tunneling events which could be due to excitations
in the electrodes or in the bath: Since the tunneling rate is very
small, these excitations will have traveled away from the double-dot
until the next event takes place. The entanglement between electron
and bath (discussed in the previous sections) would preclude factorized
initial conditions, if it were not treated indirectly in this approach
(via the unitary transformation). Note that we do not have to make
any secular approximation at this point, unlike the usual derivation
of a master equation \cite{Blum}. It turns out that all contributions
only depend on the time-difference $t_{1}-t_{2}$ anyway, because
the dot levels are degenerate. Therefore, in the long-time limit $t\rightarrow \infty $,
the integration over $(t_{1}+t_{2})/2$ results in a factor $t$,
and the endpoints of the integrals over $t_{1}-t_{2}$ may be extended
to $\infty $. This yields the desired master equation that will determine
the stationary $\hat{\rho }$, as well as the current, in the limit
of weak tunnel coupling. 

In the expectation values of products $\hat{V}'\hat{V}'$ only those
contributions remain which combine $\hat{d}_{\alpha }\hat{j}_{\alpha }$
(tunneling out of the dots) with $\hat{j}_{\beta }^{\dagger }\hat{d}_{\beta }^{\dagger }$
(tunneling onto the dots):

\begin{eqnarray}
\frac{d\hat{\rho }}{dt}=-\sum _{\alpha ,\beta ,j}\int _{0}^{\infty }ds\, \left\{ \hat{d}_{\alpha }(s)\hat{d}_{\beta }^{\dagger }\hat{\rho }\left\langle \hat{j}_{\alpha }(s)\hat{j}_{\beta }^{\dagger }\right\rangle +h.c.\right. &  & \nonumber \\
\left.+\hat{d}_{\alpha }^{\dagger }(s)\hat{d}_{\beta }\hat{\rho }\left\langle \hat{j}_{\alpha }^{\dagger }(s)\hat{j}_{\beta }\right\rangle +h.c.\right\}  &  & \nonumber \\
+\sum _{\alpha ,\beta ,j}\int _{-\infty }^{+\infty }ds\, \left\{ \hat{d}_{\alpha }(s)\hat{\rho }\hat{d}_{\beta }^{\dagger }\left\langle \hat{j}_{\beta }^{\dagger }\hat{j}_{\alpha }(s)\right\rangle +\right. &  & \nonumber \\
\left.\hat{d}_{\alpha }^{\dagger }(s)\hat{\rho }\hat{d}_{\beta }\left\langle \hat{j}_{\beta }\hat{j}_{\alpha }^{\dagger }(s)\right\rangle \right\} \, . &  & 
\end{eqnarray}
 (Note that there is no minus sign from fermion operator re-ordering
in this factorization of dot and reservoir part, as the reservoir
fermion operators are dragged past an even number of dot operators;
compare e.g. \cite{schoellerRTRG}; alternatively, it is also possible
to define them as commuting operators, since there is no interaction
between them). We get for the individual matrix elements (for brevity,
the summation over $j=l,r$ is implied):

\begin{eqnarray}
\dot{\rho }_{++} & = & -\rho _{++}\int _{-\infty }^{+\infty }ds\, \left\langle \hat{j}_{+}^{\dagger }(s)\hat{j}_{+}\right\rangle \nonumber \\
 &  & +\rho _{00}\int _{-\infty }^{+\infty }ds\, \left\langle \hat{j}_{+}\hat{j}_{+}^{\dagger }(s)\right\rangle \nonumber \\
 &  & -\rho _{-+}\int _{0}^{\infty }ds\, \left\langle \hat{j}_{+}^{\dagger }(s)\hat{j}_{-}\right\rangle -h.c.\, ,\label{DDrhoppeqGeneral}
\end{eqnarray}

\begin{eqnarray}
\dot{\rho }_{+-} & = & -\rho _{+-}\int _{0}^{\infty }ds\, \left\langle \hat{j}_{+}^{\dagger }(s)\hat{j}_{+}\right\rangle \nonumber \\
 &  & -\rho _{+-}\int _{0}^{\infty }ds\, \left\langle \hat{j}_{-}^{\dagger }\hat{j}_{-}(s)\right\rangle \nonumber \\
 &  & +\rho _{00}\int _{-\infty }^{+\infty }ds\, \left\langle \hat{j}_{-}\hat{j}_{+}^{\dagger }(s)\right\rangle \nonumber \\
 &  & -\rho _{++}\int _{0}^{\infty }ds\, \left\langle \hat{j}_{+}^{\dagger }\hat{j}_{-}(s)\right\rangle \nonumber \\
 &  & -\rho _{--}\int _{0}^{\infty }ds\, \left\langle \hat{j}_{+}^{\dagger }(s)\hat{j}_{-}\right\rangle \, .\label{DDrhopmEqGeneral}
\end{eqnarray}

The equation for $\rho _{--}$ follows from that for $\rho _{++}$
by interchanging indices $+$ and $-$. 

Now we have to evaluate environment correlators, such as the prefactor
of $\rho _{++}$ in the second equation (e.g. for $j=r$):

\begin{equation}
\left\langle \hat{r}_{+}^{\dagger }\hat{r}_{-}(s)\right\rangle =e^{i\varphi }\left\langle e^{-i\ph }e^{-i\ph (s)}\right\rangle \left\langle \hat{r}^{\dagger }\hat{r}(s)\right\rangle \, .\label{eq:rrphiphi}\end{equation}

By introducing the bare tunneling rates $\Gamma _{R(L)0}=2\pi \, D_{R(L)}\left\langle \left|t_{k}^{R(L)}\right|^{2}\right\rangle $
(compare Eq. (\ref{DDbaretunnelrate})), we get, using Eq. (\ref{DDseqrDef})
(remember $\hat{r}$ \emph{creates} a reservoir electron):

\begin{equation}
\left\langle \hat{r}^{\dagger }\hat{r}(s)\right\rangle =\frac{\Gamma _{R0}}{2\pi }\int d\epsilon \, (1-f_{R}(\epsilon ))\, e^{+i\epsilon s}\, .\end{equation}

Here we have neglected any energy-dependence of the tunnel-coupling
and electrode DOS, assuming the relevant voltages and temperatures
to be sufficiently small (but see below). The bath correlator in (\ref{eq:rrphiphi})
evaluates to $\exp (-\left\langle \ph \ph (s)\right\rangle -\left\langle \ph ^{2}\right\rangle )$,
which can be expressed by using the definition (\ref{DDPEDef}) for
$P_{-}(\omega )$. There, we have to set $s\mapsto -s$ because of
the reversed order in the $\ph $-correlator:

\begin{equation}
e^{-\left\langle \ph \ph (s)\right\rangle -\left\langle \ph ^{2}\right\rangle }=\int d\omega \, P_{-}(\omega )e^{i\omega s}\, .\end{equation}

Therefore, we obtain:

\begin{eqnarray}
\int _{0}^{\infty }ds\, \left\langle \hat{r}_{+}^{\dagger }\hat{r}_{-}(s)\right\rangle = &  & \nonumber \\
e^{i\varphi }\frac{\Gamma _{R0}}{2}\int d\epsilon \, (1-f_{R}(\epsilon ))\, \tilde{P}_{-}^{*}(-\epsilon )\, , &  & \label{DDexamplerr}
\end{eqnarray}

with

\begin{eqnarray}
\tilde{P}_{-}(\epsilon )=\frac{1}{\pi }\int d\omega \, P_{-}(\omega )\int _{0}^{\infty }ds\, e^{i(\epsilon -\omega )s}= &  & \nonumber \\
P_{-}(\epsilon )+\frac{i}{\pi }\int d\omega \, \frac{P_{-}(\omega )}{\epsilon -\omega }\, . &  & 
\end{eqnarray}

The integral in the second line is understood as a principal-value
integral. In order to abbreviate expressions like this, we introduce
the following definitions for the effective in- and out-tunneling
rates as well as the effective tunnel couplings generated by the electrodes:

\begin{eqnarray}
\gamma _{L(-)} & \equiv  & \Gamma _{L0}\int d\epsilon \, (1-f_{L}(\epsilon ))\, P_{(-)}(-\epsilon )\label{DDgammaseconddef}\\
\gamma _{L(-)}^{in} & \equiv  & \Gamma _{L0}\int d\epsilon \, f_{L}(\epsilon )\, P_{(-)}(\epsilon )\\
\Delta _{L} & \equiv  & -\frac{\Gamma _{L0}}{\pi }\int _{-\infty }^{\Lambda }d\epsilon \, (1-f_{L}(\epsilon ))\, \int d\omega \, \frac{P_{-}(\omega )}{\epsilon +\omega }\\
\tilde{\gamma }_{L-} & \equiv  & \gamma _{L-}[P\mapsto \tilde{P}]=\gamma _{L-}+i\Delta _{L}\, .\label{DDgammatildedef}
\end{eqnarray}

Analogous definitions hold for $L\mapsto R$. Eq. (\ref{DDgammaseconddef})
is equivalent to the definition (\ref{DDgammaDefinition}) used for
$\gamma _{(-)}$ in previous sections. Note that the effective tunnel
coupling $\Delta _{L(R)}$ depends on $P_{-}$, because it arises
from transitions between the states $\kt{+}$ and $\kt{-}$, via an
intermediate lead state. In the expression for $\Delta _{L(R)}$,
the energy-dependence of the density of states and the tunnel coupling
to the reservoir electrode should be kept in order to have a convergent
integral. We will take this into account by introducing an effective
upper energy cutoff $\Lambda $ in the integral. Using these definitions,
(\ref{DDexamplerr}) is equal to $\exp (i\varphi )\tilde{\gamma }_{R-}^{*}/2$.

One might wonder why the effective tunnel couplings $\Delta _{L(R)}$
do depend on the occupation of electron states in the reservoirs.
After all, in the non-interacting case, it is possible to calculate
such a change of the effective single-particle Hamiltonian prior to
filling in the electron states. Alternatively, in a calculation that
already takes into account occupation factors, there would be two
contributions which add up to an integral that does not depend on
the Fermi function. However, we consider the interacting case $U=\infty $,
such that (even without the bath) one of these contributions is missing
(since it would involve intermediate states with double occupancy).
The resulting logarithm is analogous to that which appears in the
Kondo problem. This effective tunnel coupling has also been discussed
in Ref.~\onlinecite{BoeseDD}, for the case without a bath. There, the upper
cutoff $\Lambda $ was provided by the Coulomb coupling $U$, since
for higher energies double-occupancy is no longer forbidden and the
non-interacting case takes over (where two contributions arise that
cancel each other). If we take the limit $U\rightarrow \infty $,
then $\Lambda $ will be set by a cutoff in the tunnel matrix elements
(or the electron reservoir's density of states).

The general master equation for the reduced density matrix of the
double-dot, derived in the limit of weak tunnel coupling but arbitrary
electron-bath coupling, follows by inserting the definitions (\ref{DDgammaseconddef})-(\ref{DDgammatildedef})
into Eqs. (\ref{DDrhoppeqGeneral}) and (\ref{DDrhopmEqGeneral}):

\begin{eqnarray}
\dot{\rho }_{++} & = & -\rho _{++}(\gamma _{L}+\gamma _{R})\nonumber \\
 &  & +\rho _{00}(\gamma _{L}^{in}+\gamma _{R}^{in})\nonumber \\
 &  & -\frac{\rho _{-+}}{2}(e^{i\varphi }\tilde{\gamma }_{R-}+\tilde{\gamma }_{L-})-h.c.\, ,\label{DDrhoppeq}
\end{eqnarray}

\begin{eqnarray}
\dot{\rho }_{--} & = & -\rho _{--}(\gamma _{L}+\gamma _{R})\nonumber \\
 &  & +\rho _{00}(\gamma _{L}^{in}+\gamma _{R}^{in})\nonumber \\
 &  & -\frac{\rho _{+-}}{2}(e^{-i\varphi }\tilde{\gamma }_{R-}+\tilde{\gamma }_{L-})-h.c.\, ,\label{DDrhommeq}
\end{eqnarray}

\begin{eqnarray}
\dot{\rho }_{+-} & = & -\rho _{+-}(\gamma _{L}+\gamma _{R})\nonumber \\
 &  & +\rho _{00}(e^{i\varphi }\gamma _{R-}^{in}+\gamma _{L-}^{in})\nonumber \\
 &  & -\frac{\rho _{++}}{2}(e^{i\varphi }\tilde{\gamma }_{R-}^{*}+\tilde{\gamma }_{L-}^{*})\nonumber \\
 &  & -\frac{\rho _{--}}{2}(e^{i\varphi }\tilde{\gamma }_{R-}+\tilde{\gamma }_{L-})\, .\label{DDrhopmEq}
\end{eqnarray}

The ingredients of the master equation obtained here may be interpreted
as follows: 

One part of the right hand side corresponds to the unitary time-evolution
generated by the effective tunneling Hamiltonian

\begin{equation}
\hat{H}_{eff}^{T}=\frac{1}{2}(e^{i\varphi }\Delta _{R}+\Delta _{L})\kt{+}\left\langle -\right|+h.c.\, .\end{equation}

Furthermore, the in-tunneling contributions in the equations for $\rho _{++}$
and $\rho _{--}$ depend on $P(E)$, while that for $\rho _{+-}$
is determined by $P_{-}(E)$, since it describes the creation of a
coherent superposition of $\kt{+}$ and $\kt{-}$ (which is hindered
by the bath). This term would be absent in the usual master equation.
In particular, if $\gamma _{L-}^{in}\rightarrow \gamma _{L}^{in}$,
which will be the case at $T=0$ for vanishing bias between the dots
and the left electrode, an electron tunneling from the left lead will
end up in the coherent superposition where $\rho _{+-}=\rho _{++}=\rho _{--}$.
Taking into account that we are working in a transformed basis, this
describes just the entangled state (\ref{DDentangled}), confirming
the starting point of our earlier discussion. Note that the out-tunneling
contribution for $\rho _{++}$ also depends on $\rho _{+-}$, for
example. This reflects the fact that a superposition between the two
states may be blocked from decaying into the lead, while each state
separately can decay.

The stationary density matrix is obtained by demanding $d\hat{\rho }/dt=0$
(and using the relations $\rho _{00}=1-\rho _{++}-\rho _{--}$ and
$\rho _{-+}=\rho _{+-}^{*}$). This will give us the density matrix
in zeroth order $\Gamma _{0}^{0}$ in the bare tunnel coupling, which
we need to calculate the current in leading order $\Gamma _{0}^{1}$.

We can obtain the current from the contribution of the left electrode
to the change $\dot{\rho }_{++}+\dot{\rho }_{--}$ in the double-dot
occupation (i.e. keeping only terms that stem from the left electrode
in the master equation). This is equal to the right-going current
in the stationary limit:

\begin{eqnarray}
\frac{I}{e}=(\dot{\rho }_{++}+\dot{\rho }_{--})_{L}= &  & \nonumber \\
2\rho _{00}\gamma _{L}^{in}-\gamma _{L}(\rho _{++}+\rho _{--})-2\gamma _{L-}Re[\rho _{+-}]\, . &  & \label{DDcurrentexpression}
\end{eqnarray}

An alternative way of deriving the current would be to start from
the general Meir-Wingreen formula\cite{meirwingreen} which expresses
the current in terms of the exact Green's functions of the double
dot, to be calculated in presence of the tunnel-coupling and the bath.
This has been the approach of Ref.~\onlinecite{koeniggefen} for the case
without the bath, and we have checked (\ref{DDcurrentexpression})
to give the same result in that case.

\section{Evaluation of the sequential tunneling current and the visibility}

\label{sec:Evaluation-of-the}In order to evaluate the current as
a function of temperature $T$, bias voltage $V$ and phase difference
$\varphi $, we will now specialize to the case of symmetric bias
and left-right symmetric tunnel couplings ($\Gamma _{R0}=\Gamma _{L0}=\Gamma _{0}$).
All essential features (in particular the perfect destructive interference
in absence of the bath) are independent of this assumption. We will
find that the current is symmetric under $\varphi \mapsto -\varphi $
even for the nonlinear response considered here, due to the symmetry
of the model (compare Ref.~\onlinecite{koeniggefen} for a systematic analysis
of phase-locking in a variety of interference geometries).

\begin{figure}
\begin{center}\includegraphics[  height=7cm]{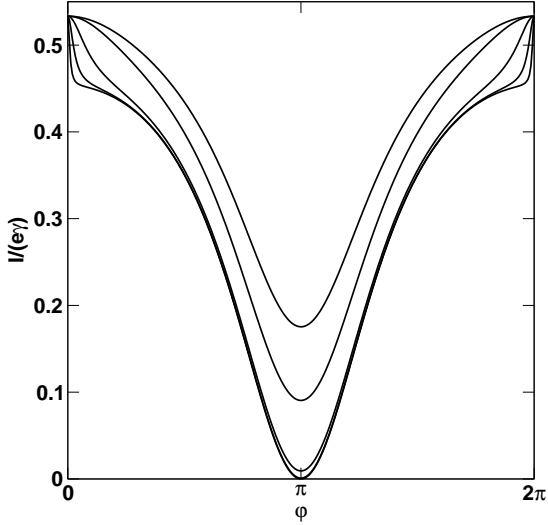}\end{center}

\caption{\label{DDIshapeFigure}The current $I$ for different values of the
visibility $\upsilon =\gamma _{-}/\gamma =0.8,\, 0.9,\, 0.99,\, 0.999,\, 0.9999$
(from top to bottom). The limits $\varphi \rightarrow 0$ and $\upsilon \rightarrow 1$
do not commute. Other parameters held fixed: $\lambda =e^{-\beta \mu }=0.2$
and $\delta _{L}=\delta _{R}=-1$. }
\end{figure}

We find from Eqs. (\ref{DDgammaseconddef})-(\ref{DDgammatildedef}),
using $f(\epsilon )=1-f(-\epsilon )$:

\begin{equation}
\gamma _{R(-)}=\gamma _{L(-)}^{in}=\gamma _{(-)}\equiv \Gamma _{0}\int d\epsilon \, f(\epsilon -\mu )\, P_{(-)}(\epsilon )\, ,\end{equation}

where $\mu =eV/2$ is the chemical potential of the left reservoir.
This is definition (\ref{DDgammaDefinition}), with $eV$ replaced
by $\mu =eV/2$ (since we deal with the symmetric bias case). Furthermore,
we use the condition of detailed balance, $P_{(-)}(-E)=\exp (-\beta E)\, P_{(-)}(E)$
(see, for example, Ref.~\onlinecite{ingoldhabil}), which leads to

\begin{equation}
\gamma _{L(-)}=\gamma _{R(-)}^{in}=e^{-\beta \mu }\gamma _{(-)}\, .\end{equation}

The effective tunnel couplings are still different (because of the
different Fermi distributions):

\begin{equation}
\Delta _{L(R)}=-\frac{\Gamma _{0}}{\pi }\int _{-\infty }^{\Lambda }d\epsilon \, f(-(\epsilon \mp \mu ))\int d\omega \, \frac{P_{-}(\omega )}{\epsilon +\omega }\, .\end{equation}

The lower sign belongs to the right electrode. 

For the special case of $T=0$, electrons always enter from the left
and go to the right, such that we have $\gamma _{L}=\gamma _{L-}=\gamma _{R}^{in}=\gamma _{R-}^{in}=0$
and $\gamma _{R(-)}=\gamma _{L(-)}^{in}=\gamma _{(-)}$, with

\begin{equation}
\gamma _{(-)}=\Gamma _{0}\int _{0}^{\mu }d\epsilon \, P_{(-)}(\epsilon )\, .\end{equation}

The effective tunnel couplings are, at $T=0$:

\begin{equation}
\Delta _{L(R)}=-\frac{\Gamma _{0}}{\pi }\int d\omega \, P_{-}(\omega )\, \ln \left[\frac{\Lambda +\omega }{|\mu \pm \omega |}\right]\, .\end{equation}

Note that, without any bath present, $\Delta _{L(R)}$ will have a
logarithmic singularity at $\mu \rightarrow 0$, for $T=0$. The upper
cutoff $\Lambda $ will be given by the minimum of the Coulomb repulsion
energy $U$ and the bandwidth of the reservoir's electronic energy
band (or by some cutoff in the tunnel matrix elements). For the purposes
of our discussion, we assume $\Lambda \gg \mu ,\omega $.

In the the limit of high bias voltages ($\omega \ll \Lambda ,\mu $),
we obtain effective tunnel couplings whose magnitude goes as $z^{2}$
and decreases logarithmically with increasing $\mu $:

\begin{equation}
\Delta _{L}\approx \Delta _{R}\approx -\frac{\Gamma _{0}}{\pi }\ln \left[\frac{\Lambda }{\mu }\right]\int d\omega \, P_{-}(\omega )\, =-z^{2}\frac{\Gamma _{0}}{\pi }\ln \left[\frac{\Lambda }{\mu }\right]\, .\end{equation}

By solving the master equation for the stationary density matrix and
inserting the result into Eq. (\ref{DDcurrentexpression}), we obtain
the expression for the current through the double dot in terms of
all of the quantities mentioned previously. In general (at arbitrary
$T$), it is found that the current may be written as the product
of $\gamma $ with a dimensionless function of the phase difference
$\varphi $ and the ratios $\upsilon =\gamma _{-}/\gamma $, $\delta _{L(R)}=\Delta _{L(R)}/\gamma $
and $\beta \mu $:

\begin{equation}
I=e\gamma \, I_{0}[\varphi ,\beta \mu ,\upsilon ,\delta _{L},\delta _{R}]\, .\end{equation}

The complete expression for $I_{0}$ is very cumbersome, although
it may be found analytically by straightforward solution of the master
equation (it is listed for $T=0$ in Appendix \ref{DDseqappendix}).
Therefore, let us first discuss the situation without coupling to
a bath. In that case, we obtain 

\begin{equation}
\delta _{L}=\delta _{R}\equiv \delta =-\frac{\Gamma _{0}}{\pi }\int _{-\infty }^{\Lambda }\frac{d\epsilon }{\epsilon }\, f(\mu -\epsilon )\, \end{equation}
 and $\gamma =\gamma _{-}=\Gamma _{0}f(-\mu )$. The current turns
out to be (with $\lambda \equiv e^{-\beta \mu }$):

\begin{equation}
\frac{I}{e\gamma }=\frac{4(1-\lambda )(\delta ^{2}+\lambda )\cos ^{2}(\frac{\varphi }{2})}{3\delta ^{2}+2(1+\lambda +\lambda ^{2})+3\delta ^{2}\cos (\varphi )}\, .\label{DDidealcurrent}\end{equation}

Several points should be noticed about this expression: Firstly, the
destructive interference at $\varphi =\pi $ remains perfect regardless
of temperature, because there are no current-carrying states at all.
At zero temperature ($\lambda =0$), the maximal amplitude of the
current is $I_{max}/e\gamma =2\delta ^{2}/(3\delta ^{2}+1)$, which
vanishes when the effective tunnel coupling $\delta $ goes to zero.
This has been explained above as a consequence of the possible transition
into a current-blocking state, which can only be undone by the effective
tunnel coupling. At finite temperatures ($\lambda >0$), the maximal
current is nonzero even for $\delta \rightarrow 0$, where it approaches
the value of $I_{max}/e\gamma =2\lambda (1-\lambda )/(1+\lambda +\lambda ^{2})$.
This has a maximum at around $T\sim \mu $. It vanishes for larger
temperatures as $\mu /T$, which is to be expected for tunneling through
a localized level (decreasing derivative of the Fermi function). In
addition, the shape of $I(\varphi )$ depends on $\delta $ and $\lambda $,
with a sharper minimum at $\varphi =\pi $ in the case of larger $|\delta |$.
In the limit of $\delta \rightarrow 0$, the current becomes a pure
cosine. At finite temperatures (as well as for $\upsilon \neq 1$)
the behaviour is similar, except for the finite amplitude of the current
at $\delta \rightarrow 0$.

Now we turn to the situation including the bath. The general expression
for the current is very lengthy, and we will omit it here. However,
it turns out that the maximal and minimal current are functions merely
of $\upsilon $ and $\lambda =e^{-\beta \mu }$, while they are independent
of $\delta _{L,R}$.

The amplitude of the minimal current (at $\varphi =\pi $) is given
by

\begin{equation}
\frac{I(\varphi =\pi )}{e\gamma }=\frac{2(1+\lambda )(1-\lambda ^{2})(1-\upsilon ^{2})}{3(1+\lambda )^{2}+(1-\lambda )^{2}\upsilon ^{2}}\, ,\end{equation}
while the maximal current (at $\varphi =0$) is

\begin{equation}
\frac{I(\varphi =0)}{e\gamma }=\frac{2}{3}(1-\lambda )\, .\label{DDmaxcurrentgeneral}\end{equation}
It should be noted that the expression (\ref{DDidealcurrent}) for
the current in the ideal case seems to contradict this simple formula.
However, that is because the limits $\varphi \rightarrow 0$ and $\upsilon \rightarrow 1$
do not commute. This is shown in Fig. \ref{DDIshapeFigure}. It means
that for $T=0$ and $\delta _{L,R}\rightarrow 0$ the maximal current
calculated according to (\ref{DDmaxcurrentgeneral}), which is independent
of $\delta _{L,R}$, and the {}``typical'' amplitude of the current
($\propto \delta _{L}^{2}$) may deviate strongly. The peculiar behaviour
near $\varphi =0$ seems to be connected to the physical degeneracy
of the case $\varphi =0,\, \upsilon =1$ which has been discussed
above.

From these formulas, we obtain the visibility, defined in terms of
the current:

\begin{equation}
\upsilon _{I}\equiv \frac{I(\varphi =0)-I(\varphi =\pi )}{I(\varphi =0)+I(\varphi =\pi )}\, .\end{equation}

It can be expressed entirely by the visibility $\upsilon $ defined
previously in terms of the tunneling rates (Eqs. (\ref{eq:visDef}),
(\ref{eq:visGamm})), as well as the temperature-dependent factor
$\lambda =e^{-\beta \mu }$ ($\mu =eV/2$):

\begin{equation}
\upsilon _{I}=\frac{2(1+\lambda +\lambda ^{2})\upsilon ^{2}}{3(1+\lambda )^{2}-(1+4\lambda +\lambda ^{2})\upsilon ^{2}}\, .\label{DDvisibilityI}\end{equation}

This is a monotonous mapping of $\upsilon $ to the interval $[0,1]$,
with only a weak dependence on $\lambda $. The other parameters $\delta _{L},\delta _{R}$
only modify the amplitude and shape of the current pattern $I(\varphi )$.
Therefore, all the statements about the visibility made in the previous
discussion of the tunneling decay out of the symmetric superposition
continue to hold up to this monotonous transformation (and with $eV$
replaced by $\mu =eV/2$). In particular, at $T=0$, we have

\begin{equation}
\upsilon _{I}=\frac{2\upsilon ^{2}}{3-\upsilon ^{2}}\, .\end{equation}

\begin{figure}
\begin{center}\includegraphics[  height=7cm]{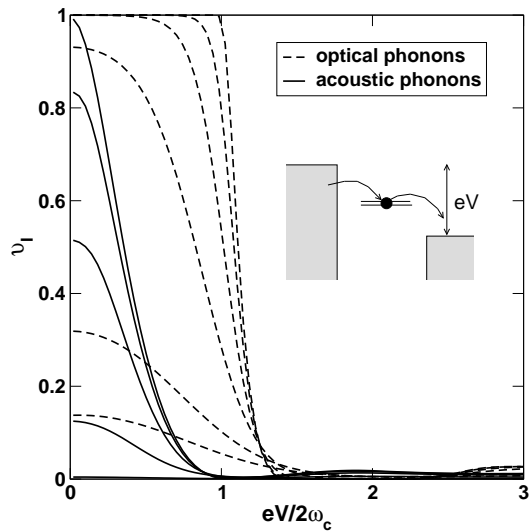}\end{center}

\caption{\label{DDvisIvsMu}The visibility $\upsilon _{I}$ of the pattern
$I(\varphi )$, for piezoelectric coupling to acoustic phonons (b)
(solid line) and for the optical phonon bath (c) (dashed line), plotted
vs. $\mu =eV/2$, at different temperatures $T/\omega _{c}=0.01,\, 0.05,\, 0.1,\, 0.2,\, 0.4,\, 0.5$
(top to bottom). Inset depicts energy diagram for tunneling in this
situation.}
\end{figure}
The dependence of the visibility $\upsilon _{I}$ on the bias voltage
$eV=2\mu $, the temperature $T$ and the bath spectrum is displayed
in Fig. \ref{DDvisIvsMu}, for bath spectra of type (b) and (c). The
decrease of $\upsilon _{I}$ at $\mu =0$ with increasing temperature
$T$ in case (b) is well approximated by Eq. (\ref{DDvisfiniteT})
for $\upsilon (T,V\rightarrow 0)$ (employing the relation $\upsilon _{I}=\upsilon ^{2}/(2-\upsilon ^{2})$
for $\mu =0$). (The functions $P_{(-)}(E)$ for finite temperatures
have been calculated numerically using the fast Fourier transform,
from the defining equation (\ref{DDPEDef})).

Note that for bath spectra with $z=0$ (i.e. exponent $s\leq 1$ at
$T=0$ and $s\leq 2$ at $T>0$) the visibility vanishes entirely
(at any $V$), as has been explained in the previous sections. We
have already pointed out that this picture is expected to change if
one treats the tunnel-coupling to higher order. However, we have to
leave this analysis for the future. One possible approach to a nonperturbative
(but still approximate) treatment of both the tunnel-coupling and
the system-bath coupling at the same time seems to be the numerical
{}``real-time renormalization group'' scheme \cite{schoellerRTRG}.

\section{Conclusions}

We have analyzed dephasing in tunneling through two parallel single-level
quantum dots with a fluctuating energy difference between the dots.
The disappearance of perfect destructive interference in a symmetric
setup has been taken as a criterion for {}``genuine'' dephasing,
as opposed to mere renormalization. The coupling to the bath has been
taken into account exactly, via the {}``independent boson model''
and the concepts of the {}``$P(E)$ theory'' of tunneling in a dissipative
environment, while the tunnel coupling has been treated in leading
order. 

We have discussed in detail the behaviour of the density matrix of
a single electron that has been placed in a superposition of the two
dot levels. The bath measures (to some extent) the position of the
electron, such that the electron's density matrix becomes mixed. However,
this allows direct conclusions about the {}``incoherent current''
only in the limit of high bias voltages, corresponding to a fast {}``projection''
measurement of the electron's state. For lower voltages, only the
low-frequency part of the bath spectrum contributes to the lifting
of destructive interference. Thus, for any {}``weak bath'', whose
spectrum falls off fast towards low frequencies, the visibility of
the interference effect becomes perfect in the limit of low bias voltages
$V$ and temperatures $T$, when the energy supplied to the electron
is vanishingly small. This is the case for a fluctuation spectrum
$\propto \omega ^{s}$ with $s>1$ ($s>2$) for $T=0$ ($T>0$). The
visibility may show a nonmonotonous behaviour as a function of bias
voltage. For {}``stronger'' spectra (smaller exponent $s$), including
the Ohmic bath ($s=1$), there is the well-known zero-bias anomaly
(suppression of the tunneling current at low voltages), which affects
equally both the cases of constructive and destructive interference.
Therefore, the visibility vanishes exactly at any bias voltage in
our approach, where the tunnel coupling has been treated only in leading
order. Although there is always a suppression of the magnitude of
the tunnel current for the case of constructive interference, this
may be interpreted as a mere renormalization of the effective tunnel-coupling,
since the perfect destructive interference is not affected and since
it occurs even for a bath with an excitation gap. The full dependence
of the sequential tunneling current $I(\varphi )$ on voltage, temperature,
bath spectrum and phase difference $\varphi $ between the interfering
paths has been derived by setting up a master equation for the state
of the double-dot (which is special due to the degeneracy of dot levels). 

The major questions that have remained open in our analysis are related
to the behaviour at stronger tunnel coupling. In particular, the perfect
destructive interference may also be overcome by correlated tunneling
of several particles (with an intermediate {}``virtual'' excitation
of the bath), and this process will therefore contribute to dephasing,
although it is expected to be suppressed strongly at low voltages
and temperatures. Likewise, the visibility for the Ohmic bath (or
other strong baths), which turns out to be zero in the present approximation,
may be changed at low bias voltages and temperatures comparable to
the tunneling rate. This will require other methods to analyze the
competition between strong tunnel coupling and system-bath coupling.

\begin{acknowledgments}
We thank J. König, H. Schoeller, D. Loss and H. Grabert for useful
discussions. This work has been supported through the Swiss NSF and
the Swiss NCCR for Nanoscience.
\end{acknowledgments}
\appendix

\section{Independent boson model}

\label{indepBosonApp}For reference purposes, we describe here the
canonical transformation employed in the independent boson model.
See Ref.\onlinecite{mahan} for more details (concerning the case
of at most a single particle). Consider a set of electronic levels
$j$ that couple to bath operators $\hat{F}_{j}$ which are assumed
to be linear in the coordinates (and momenta) of a bath of harmonic
oscillators, $\hat{H}_{B}$:

\begin{equation}
\hat{H}=\sum _{j}(\varepsilon _{j}+\hat{F}_{j})\hat{n}_{j}+\hat{H}_{B}\, .\label{indepbosonham}\end{equation}

Here $\varepsilon _{j}$ is the unperturbed level energy and $\hat{n}_{j}=\hat{d}_{j}^{\dagger }\hat{d}_{j}$
is the number of particles on level $j$. The fluctuating fields are
characterized completely by their power spectra at $T=0$, 

\begin{equation}
\left\langle \hat{F}_{l}\hat{F}_{j}\right\rangle _{\omega }^{T=0}\equiv \frac{1}{2\pi }\int _{-\infty }^{+\infty }dt\, e^{i\omega t}\, \left\langle \hat{F}_{l}(t)\hat{F}_{j}\right\rangle ^{T=0}\, .\end{equation}

Here we will restrict ourselves to the case where the different variables
commute, $[\hat{F}_{l},\hat{F}_{j}]=0$. As a consequence, the spectrum
$\left\langle \hat{F}_{l}\hat{F}_{j}\right\rangle _{\omega }^{T=0}$
is real-valued, but there may still be correlations. 

The most straightforward solution proceeds via a unitary transformation\cite{mahan}
(essentially a gauge transformation). One introduces the fluctuating
phases $\ph _{j}$, whose time-derivatives are given by the $\hat{F}_{j}$:

\begin{equation}
\dot{\ph }_{j}\equiv i[\hat{H}_{B},\ph _{j}]=-\hat{F}_{j}\, .\label{IBphidef}\end{equation}

The exponent generating the unitary transformation is defined as:

\begin{equation}
\hat{\chi }=\sum _{j}\ph _{j}\hat{n}_{j}\, .\end{equation}

Applying the transformation to the Hamiltonian in Eq. (\ref{indepbosonham})
yields:

\begin{equation}
\hat{H}'=e^{-i\hat{\chi }}\hat{H}e^{+i\hat{\chi }}=\sum _{j}\varepsilon _{j}\hat{n}_{j}-\sum _{lj}J_{lj}\hat{n}_{l}\hat{n}_{j}+\hat{H}_{B}\, .\label{IBHprime}\end{equation}

The coupling between system and bath has been eliminated, resulting
in an effective interaction between particles on the different levels,
with:

\begin{equation}
J_{lj}=\int _{0}^{\infty }d\omega \, \frac{\left\langle \hat{F}_{l}\hat{F}_{j}\right\rangle _{\omega }^{T=0}}{\omega }\, .\label{IBcoupling}\end{equation}

The $J_{lj}$ are real-valued and independent of temperature. For
$l=j$ they describe energy shifts of single-particle levels. The
canonical transformation also changes the particle annihilation and
creation operators,

\begin{equation}
\hat{d}_{j}'=e^{-i\hat{\chi }}\hat{d}_{j}e^{+i\hat{\chi }}=e^{i\ph _{j}}\hat{d}_{j}\, ,\label{IBdtransform}\end{equation}

and $\hat{d}_{j}'^{\dagger }=\hat{d}_{j}^{\dagger }e^{-i\ph _{j}}$.
This will affect all Green's functions and, therefore, also the time-evolution
of the single-particle density matrix. In addition, it becomes important
if a tunneling part is added to the Hamiltonian, where the operators
$\hat{d}_{j}^{(\dagger )}$ appear, such that they have to be transformed
according to (\ref{IBdtransform}). However, since the phases $\hat{\phi }_{j}$
and the particle operators $\hat{d}_{j}^{(\dagger )}$ commute (even
at different times, when evolved according to $\hat{H}'$), the evaluation
of Green's functions always splits into a part referring to the particles
and a separate average over the bath operators. This is the major
simplification brought about by the {}``diagonal coupling'' between
system and bath.

\section{Current expression for sequential tunneling through the double-dot}

\label{DDseqappendix}At $T=0$, for the symmetric situation, the
current $I$ is given by $I=e\gamma \, I_{0}[\upsilon ,\delta _{L},\delta _{R}]$,
with:

\begin{eqnarray}
I_{0}[\upsilon ,\delta _{L},\delta _{R}] & = & 2\cdot [-\delta _{L}^{2}+(\upsilon ^{2}-1)(1+\delta _{R}^{2})+\nonumber \\
 &  & 2\delta _{L}\delta _{R}(\upsilon ^{2}-1)\cos \varphi +\delta _{L}^{2}\upsilon ^{2}\cos ^{2}\varphi ]\cdot \nonumber \\
 &  & [-3\delta _{L}^{2}+2\delta _{L}\delta _{R}\upsilon ^{2}+(1+\delta _{R}^{2})(\upsilon ^{2}-3)+\nonumber \\
 &  & 2(\upsilon ^{2}(1+\delta _{L}^{2}+\delta _{R}^{2})+\delta _{L}\delta _{R}(\upsilon ^{2}-3))\cos \varphi +\nonumber \\
 &  & \delta _{L}(\delta _{L}+2\delta _{R})\upsilon ^{2}\cos ^{2}\varphi ]^{-1}
\end{eqnarray}

\end{document}